\documentclass[pra,showpacs,psfig,twocolumn]{revtex4}

\usepackage{amssymb}
\usepackage{graphicx}
\newtheorem{DADBN}{Theorem}
\newtheorem{DADBR}[DADBN]{Theorem}

\newtheorem{CCSP}[DADBN]{Theorem}
\newtheorem{keepRj}[DADBN]{Theorem}
\newtheorem{keepRjLem}[DADBN]{Lemma}
\newtheorem{Dby2}[DADBN]{Theorem}
\newtheorem{Dby2C}[DADBN]{Corollary}

\newtheorem{DbyN}[DADBN]{Theorem}
\newtheorem{Dbyr}[DADBN]{Corollary}
\newtheorem{R2}[DADBN]{Corollary}

\newtheorem{integer}[DADBN]{Theorem}
\newtheorem{nonint}[DADBN]{Theorem}

\begin{document}
\title{Local distinguishability with preservation of entanglement}
\author{Scott M. Cohen}
\thanks{email: cohensm@duq.edu}
\affiliation{
Department of Physics, Duquesne University, Pittsburgh, PA  15282 \\
\centerline{and} \\
Department of Physics, Carnegie-Mellon University, Pittsburgh, PA 15213
}
\date{\today}
\begin{abstract}
\noindent I consider deterministic distinguishability of a set of orthogonal, bipartite states when only a single copy is available and the parties are restricted to local operations and classical communication, but with the additional requirement that entanglement must be preserved in the process. Several general theorems aimed at characterizing sets of states with which the parties can succeed in such a task are proven. These include (1) a maximum for the number of states when the Schmidt rank of every outcome must be at least a given minimum; (2) an upper bound (equal to the dimension of Hilbert space if entanglement need not be preserved) for the sum over Schmidt ranks of the initial states when only one-way classical communication is allowed; and (3) separately, a necessary and a sufficient condition on the states such that their original Schmidt ranks can always be preserved. Two additional theorems explicitly demonstrate a tradeoff between the extent to which the set of states fill Hilbert space, as measured by their Schmidt ranks, and how refined the parties must make their measurements, an important factor in determining the Schmidt rank the state can retain after it has been identified. It is shown that our bound on the sum of Schmidt ranks can be exceeded if two-way communication is permitted, and this includes the case that entanglement need not be preserved, so that this sum can exceed the dimension of Hilbert space. Such questions, concerning how the various results are effected by the resources used by the parties (amount of classical communication and types of local operations), are addressed for each theorem. This subject is closely related to the problem of locally purifying an entangled state from a mixed state, which is of direct relevance to teleportation and dense coding using a mixed-state resource. In an appendix, I give an extremely simple and transparent proof of ``non-locality without entanglement", a phenomenon originally discussed by Bennett and co-workers several years ago.
\end{abstract}

\maketitle

\section{Introduction} Many of the most interesting problems in quantum information involve two physically separated parties each acting on their part of a shared, entangled (bipartite) state. Due to their separation, it may not be practicable to bring the parts together to perform global operations on the entire system, though the parties may have the means by which to communicate classically with each other, perhaps to share information about outcomes of measurements. Thus, the parties may perform local operations and classical communication, LOCC as it is widely known in the literature. It is a reasonable expectation that the separated parties are generally able to accomplish less with LOCC than they could if they brought the parts of the entangled system together. One example adhering to such an intuition is that only if the parts are brought together is it possible to increase the entanglement of the system on average \cite{BennettMixedQEC}.

As is often the case in discussions of entangled systems, however, there have been surprises in store for us. One such surprise was provided by Walgate and co-workers \cite{Walgate}, who showed that it is not necessary to use global operations in order to distinguish between two orthogonal, multipartite pure states. That is, if the parties are given a system that is in one of two possible orthogonal states, they can with certainty determine which state their shared system is in by means of LOCC alone. If the two states to be distinguished are product states, with an absence of quantum correlations between the parts, there is of course no reason to have expected otherwise. However, when the parts are correlated through quantum entanglement, one might have expected a need for global operations to learn about the differing correlations present in the two  states to be distinguished. 

To accomplish the task of distinguishing, it was shown in \cite{Walgate} that for any two orthogonal states, the parties simply need perform standard, projective measurements on their separate parts, with one party's measurement conditioned on the other party's outcome, communicated to the former by means of a classical channel. By standard, here, I mean that the measurements involve projections onto one-dimensional subspaces (pure states) of the Hilbert space describing states of each local system. Thus, although the parties have succeeded in determining the state they \textit{were} given, the state they \textit{now} share is, with certainty, a product state. Hence, this means of determining the state leads, necessarily, to a concomitant destruction of entanglement.

It is by now well understood that entanglement is an important resource, examples provided by its use in quantum communication protocols such as teleportation \cite{BennettTele} and dense coding \cite{BennettDense}. Therefore, along with a desire by the two separated parties to discover which of the two (or more) states they share, they may also wish to preserve at least a portion of the entanglement inherent in those states. For example, suppose that Alice wishes to use teleportation to convey to Bob quantum information in the form of a quantum state. Suppose also that they share an entangled two-qubit system, described by a rank-2 mixed state, such that the two qubits may be viewed as being in one of two orthogonal maximally entangled states, but it is not known which one of these states they are in. Before teleportation may be accomplished, the parties must determine which state actually describes their pair of qubits.

For the above example of entangled qubits, we will see that it is in fact not possible to determine which state the qubits are in, while at the same time preserving entanglement. However, for higher-dimensional systems, there are conditions under which both tasks may be accomplished simultaneously. This is the question we wish to study: When can a set of orthogonal, bipartite states be distinguished by LOCC while preserving some part of the original entanglement present in those states? In the next section, I will argue that Schmidt ranks of the states provide a useful characterization of these sets, as well as insight into ways of thinking about this problem. Most of our results will be stated in these terms.

Throughout the paper, consideration will be restricted to cases where there is only a single copy of the given state available to the parties. An LOCC protocol should be understood to mean that the parties perform a sequence of measurements with each outcome communicated to the other party, who may then use that information in choosing the next measurement in the sequence. The final outcome is represented by an operator $A\otimes B$, with $A$ and $B$ each equal to (ordered) products of Kraus operators \cite{Kraus} $\{A_l^{(\mu)}\}$ or $\{B_m^{(\nu)}\}$ corresponding to individual outcomes ($l,m$) in the sequence of measurements (here labeled by $\mu,\nu$). 

In this paper, we will mainly be concerned with deterministic distinguishing, whereby the parties are always able to determine which state they have. Unless explicitly indicated otherwise, the term ``distinguishing'' should be understood in this sense. In this case, the requirement of completeness may be imposed; for example,
\begin{equation}\label{complete}
	\sum_{l} A_l^{(\mu)\dagger} A_l^{(\mu)}=I_A,
\end{equation}
with $I_A$ the identity operator on Alice's Hilbert space ${\cal H}_A$. By the statement that the state $|\Psi_j\rangle$ is identified (or distinguished) by outcome $A\otimes B$ while preserving Schmidt rank $r_j$, we will mean that the Schmidt rank of the residual state $(A\otimes B)|\Psi_j\rangle$ is $r_j$, and that $(A\otimes B)|\Psi_k\rangle = 0,~\forall_{k\ne j}$. 

The paper is organized as follows: In the next section, a simple example is given to illustrate the basic ideas, and then it is argued that Schmidt ranks will be a useful quantity for characterizing distinguishability with preservation of entanglement. In Section~\ref{protocols}, a brief outline of the types of protocols to be discussed is given. Section~\ref{main} presents the main theorems including (1) a maximum for the number of states in the case that every outcome must preserve a fixed minimum Schmidt rank; (2) an upper bound for the sum over Schmidt ranks of the initial states when only one-way classical communication is allowed, again assuming that every outcome preserves a fixed minimum Schmidt rank; and (3) conditions on the states such that their original Schmidt ranks can always be preserved. Two additional theorems are given in Section~\ref{resRank}, explicitly demonstrating a tradeoff between the extent to which the set of states fill Hilbert space, as measured by their Schmidt ranks, and how refined the parties must make their measurements. In all cases, the effects of restricting or expanding the resources available to the parties (types of local operations and amount of classical communication) is discussed. In particular, various examples are given where two-way communication allows the parties to accomplish what our theorems show cannot be accomplished with one-way communication alone. One of these examples demonstrates that the sum of Schmidt ranks can exceed the dimension of Hilbert space, yet the states can nonetheless be deterministically distinguished. Other examples show that when general separable operations are allowed, the parties can do even better than they can using LOCC with two-way communication. Several of the simpler proofs of these theorems are included in this section, whereas the more lengthy proofs are given in Appendix~\ref{proofs}. Then, in Section~\ref{disc}, I point out the close correspondence of the present study to the important question of using LOCC to obtain a pure entangled state from a single copy of a mixed state. Extension to multipartite systems for two of the theorems is also discussed. Finally, in Section~\ref{sum}, I present a summary of the results. In one of the appendices, I include a very simple proof of nonlocality without entanglement, using a transparent and intuitively clear argument. The final appendix discusses LOCC protocols where the parties are not allowed to communicate until after they have completed their measurements.

\section{Characterization by Schmidt ranks}
\label{ex}

In this section I argue that it will be useful, in characterizing a set of states to be distinguished, to consider the Schmidt ranks, $R_j$, of the states in that set. Let us begin with a very simple example to illustrate the general idea of distinguishing and preserving entanglement (additional examples will appear in the following sections as illustrations of the theorems). The example involves two states on a $4 \times 4$ system (I omit normalization where it is unimportant): 
\begin{eqnarray}
	\label{exstates}
	|\Psi_1\rangle = |02\rangle_{AB} + |13\rangle_{AB} + |20\rangle_{AB} + |31\rangle_{AB}, \nonumber \\
	|\Psi_2\rangle = |00\rangle_{AB} + |11\rangle_{AB} + |22\rangle_{AB} + |33\rangle_{AB}.
\end{eqnarray}
Alice and Bob perform measurements, each with two outcomes corresponding to projectors, which with $\alpha = A$ or $B$, are
\begin{eqnarray}
	P_{\alpha 1} & = & |0\rangle_{\alpha }\langle 0| + |1\rangle_{\alpha }\langle 1|, \nonumber \\
	P_{\alpha 2} & = & |2\rangle_{\alpha }\langle 2| + |3\rangle_{\alpha }\langle 3|.
\end{eqnarray}
If Alice obtains outcome $2$, for example, then
\begin{eqnarray}
	|\widetilde\Psi_1\rangle = P_{A 2}|\Psi_1\rangle = |20\rangle_{AB} + |31\rangle_{AB}, \nonumber \\
	|\widetilde\Psi_2\rangle = P_{A 2}|\Psi_2\rangle = |22\rangle_{AB} + |33\rangle_{AB},
\end{eqnarray}
\noindent which leaves Bob with reduced density operators ($\widetilde\rho_j^B = \textrm{Tr}_{A}(|\widetilde\Psi_j\rangle\langle\widetilde\Psi_j|)$) proportional to $P_{B1}$ or $P_{B2}$, respectively. Then if he obtains outcome $1$, and if they communicate their results to each other, they will know that the state was $|\Psi_1\rangle$, and more importantly, that they now share the state $|\widetilde\Psi_1\rangle$.
Any other pair of outcomes leads to the same sort of conclusion: they know which state they had, and also know the state that remains, that being uniformly entangled across $2$-dimensional subspaces. In this example, Bob was able to use the same measurement regardless of Alice's outcome, though they still had to communicate classically in order to determine the state.

\begin{figure}
\centering
\begin{picture}(160,100)
	\put(-24,92){\makebox(24,20){$|0\rangle_B$}\makebox(24,20){$|1\rangle_B$}\makebox(24,20){$|2\rangle_B$}\makebox(24,20){$|3\rangle_B$}}
	\put(-48,12){\makebox(20,20){$|3\rangle_A$}}
	\put(-48,32){\makebox(20,20){$|2\rangle_A$}}
	\put(-48,52){\makebox(20,20){$|1\rangle_A$}}
	\put(-48,72){\makebox(20,20){$|0\rangle_A$}}
	{\thicklines\put(-24,12){\framebox(96,80)}} 
	\put(-24,12){\framebox(24,20){}\framebox(24,20){1}\framebox(24,20){}\framebox(24,20){2}} 
	\put(-24,32){\framebox(24,20){1}\framebox(24,20){}\framebox(24,20){2}\framebox(24,20){}} 
	\put(-24,52){\framebox(24,20){}\framebox(24,20){2}\framebox(24,20){}\framebox(24,20){1}} 
	\put(-24,72){\framebox(24,20){2}\framebox(24,20){}\framebox(24,20){1}\framebox(24,20){}} 
	\put(12,-4){\makebox(24,8){(a)}}

	\put(104,92){\makebox(24,20){$|0^\prime\rangle_B$}\makebox(24,20){$|1^\prime\rangle_B$}\makebox(24,20){$|2^\prime\rangle_B$}\makebox(24,20){$|3^\prime\rangle_B$}}
	\put(80,12){\makebox(20,20){$|3^\prime\rangle_A$}}
	\put(80,32){\makebox(20,20){$|2^\prime\rangle_A$}}
	\put(80,52){\makebox(20,20){$|1^\prime\rangle_A$}}
	\put(80,72){\makebox(20,20){$|0^\prime\rangle_A$}}
	{\thicklines\put(104,12){\framebox(96,80)}} 
	\put(104,12){\framebox(24,20){1,2}\framebox(24,20){1,2}\framebox(24,20){1,2}\framebox(24,20){1,2}} 
	\put(104,32){\framebox(24,20){1,2}\framebox(24,20){1,2}\framebox(24,20){1,2}\framebox(24,20){1,2}} 
	\put(104,52){\framebox(24,20){1,2}\framebox(24,20){1,2}\framebox(24,20){1,2}\framebox(24,20){1,2}} 
	\put(104,72){\framebox(24,20){1,2}\framebox(24,20){1,2}\framebox(24,20){1,2}\framebox(24,20){1,2}} 
	\put(140,-4){\makebox(24,8){(b)}}
\end{picture}
\caption{\label{fig:ex}Representation of the set of states given in Eq.~(\ref{exstates}). Alice's basis states are denoted along the left side of each grid; Bob's along the top. The numbers ($j$) inside the boxes indicate the state ($|\Psi_j\rangle$) has the corresponding product state as a component. (a) For the bases of Eq.~(\ref{exstates}), it is easily seen that the states can be distinguished by LOCC, preserving Schmidt rank of $2$ for all outcomes. (b) When viewed in other bases, the distinguishability may be far from obvious.}
\end{figure}

The representation of the original states in Fig.~\ref{fig:ex}(a) provides intuition as to what can and cannot be accomplished. It is apparent from this diagram that the two states each fill too much of the space, and are too intertwined \cite{twined} with each other, for it to be possible to distinguish and preserve Schmidt rank of $r=4$ even for a single (LOCC) outcome. On the other hand, it is quite clear that there is ``room" enough for them to be distinguished preserving $r=2$. Of course, since we are dealing with quantum systems, an unlimited number of other bases are available to us for representing these states. One other choice is shown in Fig.~\ref{fig:ex}(b), in which the distinguishability of the states is by no means clear, let alone the possibility of preserving entanglement in the process. In the latter depiction, it appears that each state by itself fills the whole space. If we were dealing with classical probability distributions, this conclusion would be correct and distinguishing the distributions would be impossible. For the case we are considering, however, the existence of quantum superpositions forces us to re-examine what is meant by the notion of ``filling space". We need a way to measure how much space a given state occupies, and if possible, to what extent the states are intertwined with each other. The Schmidt rank of the states provides such a description. When a state is written in its Schmidt basis, the part of the space it ``occupies" is minimized, and it is apparent in diagrams such as Fig.~\ref{fig:ex}(a) just where that region is. Furthermore, one can see in this diagram the level of intertwining amongst the states, and this would be true, at least qualitatively, even if the bases used were the Schmidt bases for only one of the states. This argument should at least make it plausible that consideration of Schmidt ranks, $R_j$, of the original states will be an advantageous approach, and this is what we shall do in the following sections. We shall also find it useful to consider Schmidt ranks, $r_k$, of residual states, those remaining after the parties complete their measurements. We will see in the various theorems below, that the ranks $R_j$ and the amount of entanglement that can be preserved, as measured in somewhat qualitative terms by $r_k$, are two closely related quantities.

The protocol described above succeeds in preserving entanglement by partitioning the respective spaces into subspaces that are larger than one-dimensional. In contrast, the approach of \cite{Walgate} utilized projections onto pure states, which is clearly inadequate for the purpose of preserving entanglement since it always leaves them with a product state. One way of looking at this is that they have constricted the states too much, squeezing out all of the entanglement. This difficulty can be overcome by relaxing one's grip, projecting onto higher-dimensional subspaces in making measurements. The tradeoff is that more entanglement means less information: the higher the dimensionality of the subspaces in the partitions, the more entanglement can be preserved, but less information about the state is obtained, making it more difficult to distinguish the states. So while it is always possible to distinguish a pair of orthogonal states, the added requirement of preserving entanglement leads us to a very rich structure with many challenging and interesting problems to investigate. We begin such a study in the following sections.

\section{Types of Protocols}
\label{protocols}

Our aim in this paper is to characterize sets of bipartite states, ${\cal S} = \{|\Psi_j\rangle\}_{j=1}^N$, which allow the parties to distinguish while preserving entanglement. Such a characterization does not depend solely on properties of $\cal S$, however, but also on the tools that are available to the parties as they attempt to accomplish this task. In general, we will restrict the parties to LOCC, so there are two such tools we wish to consider: (1) the types of local operations (LO) they are able to implement; and (2) the amount of classical communication (CC) they are allowed to share with each other. Whenever possible, extension to the class of separable operations \cite{Rains} will be considered, with comments on how this might enlarge the class of allowable sets $\cal S$. 

The two types of local operations to be discussed are orthogonal, projective measurements and generalized measurements. As illustration, suppose Alice performs a measurement represented by the set of Kraus \cite{Kraus} operators $A_l$, which must obey the completeness relation, Eq.~(\ref{complete}). For a generalized measurement, this is the only constraint, while for an orthogonal, projective measurement, we also have that $A_lA_{l^\prime} = A_l\delta_{ll^\prime}$. In the latter case, the local Hilbert space is divided into mutually orthogonal subspaces by the operators $A_l$, whereas in the general case the $A_l$ may divide the space into subspaces that overlap with each other to an arbitrary extent. For classical communication, we will assume the parties have access to a classical channel that can either carry information in only one direction (one-way CC) or in both directions (two-way CC). In the former case, one party must measure first and then communicate the outcome of their measurement to the other party, who must then complete the protocol without additional assistance from the first party. For two-way CC, they can go back-and-forth measuring and exchanging information as many times as is needed, conditioning subsequent measurements on previous outcomes. We will also consider protocols where the parties are only allowed to communicate after they have completed their measurements.

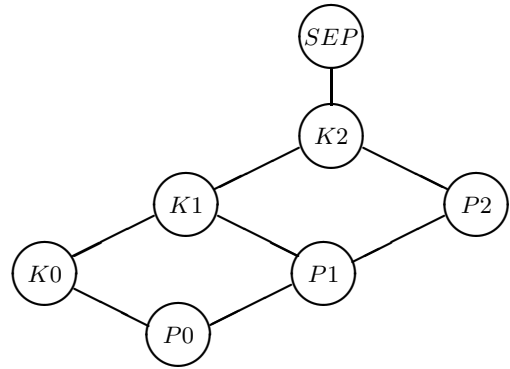
\begin{figure}
\centering
\begin{picture}(180,120)
\setlength{\unitlength}{0.04in}
\thicklines
  \label{ket}
  \put(40,40){\circle{8}}
  \put(40,40){\makebox(0,0){$\small{SEP}$}}
  \put(40,36){\line(0,-1){5}}
  \put(40,27){\circle{8}}
  \put(40,27){\makebox(0,0){$\small{K2}$}}
  \put(35.8,25.5){\line(-2,-1){11}}
  \put(44.2,25.5){\line(2,-1){11}}
  \put(59,18){\circle{8}}
  \put(59,18){\makebox(0,0){$\small{P2}$}}
  \put(54.8,16.5){\line(-2,-1){12}}
  \put(21,18){\circle{8}}
  \put(21,18){\makebox(0,0){$\small{K1}$}}
  \put(16.8,16.5){\line(-2,-1){10.6}}
  \put(25.2,16.5){\line(2,-1){10.5}}
  \put(39,9){\circle{8}}
  \put(39,9){\makebox(0,0){$\small{P1}$}}
  \put(34.8,7.5){\line(-2,-1){10.6}}
  \put(2.5,9){\circle{8}}
  \put(2.5,9){\makebox(0,0){$\small{K0}$}}
  \put(6.4,7.0){\line(2,-1){9.7}}
  \put(20,1){\circle{8}}
  \put(20,1){\makebox(0,0){$\small{P0}$}}
\end{picture}
\caption{\label{fig:lttc}Lattice diagram showing relationships between the various types of protocols discussed in the text. A connected path upward (downward) from one vertex, possibly passing through others, to a second vertex indicates that a sufficient (necessary) condition on the first implies the same for the second.}
\end{figure}

Thus, we consider seven types of protocols:
\begin{enumerate}
	\item Orthogonal projectors with CC only after measurements are completed (LOCC-P0)
	\item Generalized (Kraus) operations with CC only after measurements are completed (LOCC-K0)
	\item Orthogonal projectors with one-way CC (LOCC-P1)
	\item Generalized (Kraus) operations with one-way CC (LOCC-K1)
	\item Orthogonal projectors with two-way CC (LOCC-P2)
	\item Generalized (Kraus) operations with two-way CC (LOCC-K2)
	\item Separable operations (SEP)
\end{enumerate}
Since a projector is a Kraus operator (but not vice-versa), and since LOCC operations are a proper subset of SEP, there is a trend toward more general protocols as one moves down the list, as well as toward more sophistication in the resources needed to implement them. Therefore, conditions on $\cal S$ necessary and/or sufficient for one of these types of protocols have implications for other protocol types. Figure~\ref{fig:lttc} illustrates the specific relationships. 
We will be interested in characterizing the sets of states that allow distinguishing with preservation of entanglement for each of the protocol types. Ideally, one would like to have a complete characterization describing precisely which sets of states are allowable in each case. Such a lofty goal must await further efforts, but I hope, nonetheless, that the results presented below will be of some interest to the reader. 

\section{Main Theorems}
\label{main}
Our main results are presented in this section, in the form of several theorems and corollaries aimed at characterizing sets of states which allow two parties to distinguish while preserving entanglement using LOCC. These results are discussed with particular attention to how these characterizations may be effected by a change in the specific type of LOCC (or SEP) protocol used. For example, if a theorem provides a bound on a certain quantity, such as the number of states that can be included in the set, that bound, as well as whether or not it is a tight bound, may depend on the protocol. Whenever possible, we provide comments on such questions. To aid the flow of the discussion, the longer proofs are not presented here, but are given in an appendix.

\subsection{Maximum number of states}
\label{nmax}

If the parties need not preserve entanglement, there is an obvious upper bound, $N \le D_AD_B$, on the number of states that can be distinguished by LOCC (or otherwise) if the space has dimension $D_AD_B$. The following theorem generalizes this result to the case where entanglement must be preserved, providing a relationship between the filling of Hilbert space by the initial states and the entanglement (Schmidt ranks $r_k$) that can be preserved. 

\begin{DADBN} \label{DADBn} Suppose the parties share a $D_A \times D_B$ system and using LOCC-K2 are able to distinguish with certainty amongst a set of $N$ states while preserving Schmidt rank of at least $r$ for every outcome. Then, 
\begin{equation}
	N \le \lfloor D_A/r\rfloor \lfloor D_B/r\rfloor \equiv N_{\max},
\end{equation}
where $\lfloor x\rfloor$ is the largest integer not greater than $x$, and this upper bound is achievable by LOCC-P0.
\end{DADBN}

Thus, for each of the types of LOCC, the number of states can be as large as $N_{\max}$ and no larger. The idea of the proof, presented in Appendix \ref{nmax_proof}, is to sequentially introduce divisions of Hilbert space into orthogonal subspaces in a way consistent with LOCC and such that after the final division, no subspace has dimension larger than $r$ on either party's side. The maximum number of states on the whole space is then bounded above by the sum over maximum numbers of states on the subspaces, which for this particular method of division yields the upper bound given in the theorem. This bound is tight for all (LOCC) protocol types, as there exist sets of $N_{\max}$ states that can be distinguished by LOCC-P0. Such a set of states is depicted in Fig.~\ref{fig:nmax}, where each of the numbered square blocks represents an $r\times r$ subspace.

\begin{figure}
\centering
\begin{picture}(64,100)
	{\thicklines\put(-30,0){\framebox(132,110)}} 
	\put(-30,90){\framebox(24,20){1}\framebox(24,20){2}\framebox(24,20){3}\framebox(24,20){4}\framebox(24,20){5}\framebox(12,20){}} 
	\put(-30,70){\framebox(24,20){6}\framebox(24,20){7}\framebox(24,20){8}\framebox(24,20){9}\framebox(24,20){10}\framebox(12,20){}} 
	\put(-30,50){\framebox(24,20){11}\framebox(24,20){12}\framebox(24,20){13}\framebox(24,20){14}\framebox(24,20){15}\framebox(12,20){}} 
	\put(-30,30){\framebox(24,20){16}\framebox(24,20){17}\framebox(24,20){18}\framebox(24,20){19}\framebox(24,20){20}\framebox(12,20){}} 
	\put(-30,10){\framebox(24,20){21}\framebox(24,20){22}\framebox(24,20){23}\framebox(24,20){24}\framebox(24,20){25}\framebox(12,20){}} 
	\put(-30,0){\framebox(24,10){}\framebox(24,10){}\framebox(24,10){}\framebox(24,10){}\framebox(24,10){}\framebox(12,10){}} 
\end{picture}
\caption{\label{fig:nmax}A set of states achieving the bound of Theorem~\ref{DADBn} using LOCC-P0. Each numbered box represents an $r\times r$ subspace, and $N_{\max}=25$ with $\lfloor D_A/r\rfloor = 5 = \lfloor D_B/r\rfloor$ in this example.}
\end{figure}

If separable measurements are used, it is possible to have $N>N_{\max}$. A specific example \cite{lyu} of such a set in $3\times3$ with $r=2$ is
\begin{eqnarray}\label{eqn:Yu}
	|\Psi_1\rangle = |00\rangle + |22\rangle,\nonumber\\
	|\Psi_2\rangle = |01\rangle + |12\rangle,\nonumber\\
	|\Psi_3\rangle = |10\rangle + |21\rangle.
\end{eqnarray}
The separable POVM ($E_{mn} = A_{mn}^\dagger A_{mn}\otimes B_{mn}^\dagger B_{mn}$, with $A_{mn}$, $B_{mn}$ the corresponding Kraus operators) which distinguishes this set is
\begin{eqnarray}\label{eqn:POVM}
	E_{11} = \alpha(|0\rangle_A\langle0| + \beta|2\rangle_A\langle2|)\otimes(|0\rangle_B\langle0| + \beta|2\rangle_B\langle2|),\nonumber\\
	E_{12} = \alpha(\beta|0\rangle_A\langle0| + |2\rangle_A\langle2|)\otimes(\beta|0\rangle_B\langle0| + |2\rangle_B\langle2|),\nonumber\\
	E_{21} =\alpha(|0\rangle_A\langle0| + \beta|1\rangle_A\langle1|)\otimes(|1\rangle_B\langle1| + \beta|2\rangle_B\langle2|),\nonumber\\
	E_{22} =\alpha(\beta|0\rangle_A\langle0| + |1\rangle_A\langle1|)\otimes(\beta|1\rangle_B\langle1| + |2\rangle_B\langle2|),\nonumber\\
	E_{31} = \alpha(|1\rangle_A\langle1| + \beta|2\rangle_A\langle2|)\otimes(|0\rangle_B\langle0| + \beta|1\rangle_B\langle1|),\nonumber\\
	E_{32} = \alpha(\beta|1\rangle_A\langle1| + |2\rangle_A\langle2|)\otimes(\beta|0\rangle_B\langle0| + |1\rangle_B\langle1|),
\end{eqnarray}
with $\alpha = (2-\sqrt{3})/4$, $\beta = 2+\sqrt{3}$, and $\sum_{m,n}E_{mn} = I_A\otimes I_B$. $E_{mn}$ identifies state $|\Psi_m\rangle$ and preserves $r=2$ in all cases. Although $N_{\max} = \lfloor 3/2\rfloor\lfloor 3/2\rfloor = 1$, we here have three states in a set that is deterministically distinguished by SEP preserving $r=2$. These states are depicted in Fig.~\ref{fig:Yu}. It should be at least plausible from this diagram that no LOCC protocol can succeed at this task, and that this remains true even if one of the three states is removed, a view confirmed by the theorem.

\begin{figure}
\centering
\begin{picture}(60,80)
	\put(0,60){\makebox(24,20){$|0\rangle_B$}\makebox(24,20){$|1\rangle_B$}\makebox(24,20){$|2\rangle_B$}}
	\put(-24,40){\makebox(20,20){$|0\rangle_A$}}
	\put(-24,20){\makebox(20,20){$|1\rangle_A$}}
	\put(-24,0){\makebox(20,20){$|2\rangle_A$}}
	{\thicklines\put(0,0){\framebox(72,60)}} 
	\put(0,40){\framebox(24,20){1}\framebox(24,20){2}\framebox(24,20){}} 
	\put(0,20){\framebox(24,20){3}\framebox(24,20){}\framebox(24,20){2}} 
	\put(0,0){\framebox(24,20){}\framebox(24,20){3}\framebox(24,20){1}} 
\end{picture}
\caption{\label{fig:Yu}The three states of Eq.~\ref{eqn:Yu}, which can be distinguished by the SEP POVM of Eq.~\ref{eqn:POVM} preserving Schmidt rank of $r=2$ for all outcomes, a task that cannot be accomplished by any LOCC protocol.}
\vspace{.05in}
\end{figure}

\subsection{Schmidt rank sum for one-way protocols}\label{rankSum}
\subsubsection{An upper bound for one-way CC}

We now consider the sum over Schmidt ranks ($R_j$) of a set of states ($|\Psi_j\rangle$) which is perfectly distinguishable by LOCC on $D_A\times D_B$. There exist a number of interesting results in the literature \cite{Ghosh,GhoshPRL,Ghosh2,Horodecki2Sen2,Nathanson} that together suggest the following intuitively pleasing upper bound on this sum: $\sum_{j} R_j \le D_A D_B $ (there is no consideration of preserving entanglement in these papers). For example, (1) no more than $D$ maximally entangled states on a $D\times D$ system ($\sum R_j \le D^2$) can be perfectly distinguished \cite{Nathanson} (see also \cite{Ghosh2,GhoshPRL}), and (2) if a complete basis is perfectly distinguishable, it must be a product basis \cite{Horodecki2Sen2}. In these papers, the parties are allowed to use two-way communication. The following theorem generalizes this upper bound to the case where entanglement must be preserved, but is proved only for a restriction to one-way classical communication. In fact, I will show by means of counter-examples in the next subsection that the bound in this theorem can be exceeded when two-way communication is allowed, and this includes the case that entanglement need not be preserved ($r=1$ in the theorem).

\begin{DADBR} \label{DADBr} If Alice goes first using LOCC-K1 and the parties are always able to distinguish and preserve Schmidt rank at least $r$, then $\sum_{j} R_j \le D_A \lfloor D_B/r\rfloor$.
\end{DADBR}
The proof is given in Appendix~\ref{sec:Schmidt_proof}. Another rather obvious upper bound is $D_{\min}N_{\max}$, where $N_{\max}$ is given in Theorem~\ref{DADBn} and $D_{\min}$ is the smaller of $D_A$ and $D_B$. Clearly, the bound in the theorem cannot be achieved if it is larger than $D_{\min}N_{\max}$. Assuming both dimensions are at least $r$ (otherwise $N_{\max}=0$), this will only be the case when $D_{\min} = D_B < D_A/\lfloor D_A/r\rfloor < 2r$ so that $N_{\max} = \lfloor D_A/r\rfloor$. In this case, $D_B\lfloor D_A/r\rfloor$ is a tight upper bound, realized by a set of $\lfloor D_A/r\rfloor$ rank-$D_B$ states placed into orthogonal $D_B\times D_B$ subspaces. Then, the parties can always preserve $D_B\ge r$ using LOCC-P0 (Bob need not measure at all). For all other cases, the bound in the theorem is tight (for LOCC-K1), as is also shown in Appendix~\ref{sec:Schmidt_proof}.

There are many cases where the bound of Theorem~\ref{DADBr} can be reached when the parties can only use LOCC-P0. For example, if $D_A$ is divisible by $r$, it is easy to construct a set of states that will do this, such as can be visualized by deleting the small rectangles along the bottom of Figure~\ref{fig:nmax}. Nonetheless, there are also cases where LOCC-P1 is not sufficient for the parties to succeed unless the Schmidt rank sum is strictly less than $D_A\lfloor D_B/r\rfloor$, and we do not have a tight bound that applies in general for this type of protocol. An example is $D_A = 2r+1 = D_B$, where the bound in the theorem is $4r+2$. Suppose the set contains the maximum of $4$ states (other cases may be analyzed in a similar way). The best Alice can do with orthogonal projectors is to divide her space into two subspaces, one of dimension $r$ and the second of dimension $r+1$, and the same goes for Bob after he is informed of her outcome. When Alice obtains the $r$-dimensional outcome, the two states left (if Bob can then distinguish preserving $r$) must be Schmidt rank $R_j=r$. If one of them has rank greater than this, part of that state will lie in Alice's other subspace, meaning that when she obtains her $(r+1)$-dimensional outcome, they will not be able to distinguish the other pair of states from this one. On the other hand, when Alice gets her larger outcome, one of the second pair of states could have had rank $R_j=r+1$, but the other must have had rank $r$ or else it will not now be distinguishable from the first (to see this apply the theorem, noting that now Bob is going first). Hence, three of the states must have started with rank-$r$, the fourth with rank $r+1$, and the sum of these ranks is $4r+1<4r+2$; the bound cannot be achieved.

\subsubsection{Doing better with two-way CC}
\label{2-way}
We will now see that the upper bound in Theorem~\ref{DADBr} does not apply if two-way communication is allowed. An example is given by the four states on a $5\times 5$ system represented in Figure~\ref{fig:beatSchmidt}, with
\begin{eqnarray}\label{beatRj}
	|\Psi_1\rangle & = & |00\rangle_{AB} + |11\rangle_{AB}, \nonumber \\
	|\Psi_2\rangle & = & |02\rangle_{AB} + |13\rangle_{AB} + |24\rangle_{AB},\nonumber \\
	|\Psi_3\rangle & = & |20\rangle_{AB} + |31\rangle_{AB} + |42\rangle_{AB}, \nonumber \\
	|\Psi_4\rangle & = & |04\rangle_{AB} + |22\rangle_{AB} + |33\rangle_{AB} + |40\rangle_{AB}.
\end{eqnarray}
The sum of Schmidt ranks is now $12 > D_A\lfloor D_B/r\rfloor = 10$, with $r = 2$. Alice starts with the following pair of measurement operators,
\begin{eqnarray}
	A_{1} & = & |0\rangle_{A }\langle 0| + |1\rangle_{A}\langle 1| + |2\rangle_{A}\langle 2| + \frac{1}{\sqrt{2}}|3\rangle_{A}\langle 3|, \nonumber \\
	A_{2} & = & \frac{1}{\sqrt{2}}|3\rangle_{A}\langle 3| + |4\rangle_{A}\langle 4|.
\end{eqnarray}
If she gets outcome $A_1$, Bob designs his measurement as
\begin{eqnarray}
	B_{1} & = & |0\rangle_{B}\langle 0| + |1\rangle_{B}\langle 1|, \nonumber \\
	B_{2} & = & |2\rangle_{B}\langle 2| + \frac{1}{\sqrt{2}}|3\rangle_{B}\langle 3|, \nonumber \\
	B_{3} & = & \frac{1}{\sqrt{2}}|3\rangle_{B}\langle 3| + |4\rangle_{B}\langle 4|,
\end{eqnarray}
after which Alice can then distinguish and preserve $r=2$ in all cases. If Alice gets outcome $A_2$, then Bob can easily distinguish the remaining states and again preserve $r=2$.

\begin{figure}
\centering
\begin{picture}(64,120)
	\put(-10,100){\makebox(24,20){$|0\rangle_B$}\makebox(24,20){$|1\rangle_B$}\makebox(24,20){$|2\rangle_B$}\makebox(24,20){$|3\rangle_B$}\makebox(24,20){$|4\rangle_B$}}
	\put(-34,0){\makebox(20,20){$|4\rangle_A$}}
	\put(-34,20){\makebox(20,20){$|3\rangle_A$}}
	\put(-34,40){\makebox(20,20){$|2\rangle_A$}}
	\put(-34,60){\makebox(20,20){$|1\rangle_A$}}
	\put(-34,80){\makebox(20,20){$|0\rangle_A$}}
	{\thicklines\put(-10,0){\framebox(120,100)}} 
	\put(-10,80){\framebox(24,20){1}\framebox(24,20){}\framebox(24,20){2}\framebox(24,20){}\framebox(24,20){4}} 
	\put(-10,60){\framebox(24,20){}\framebox(24,20){1}\framebox(24,20){}\framebox(24,20){2}\framebox(24,20){}} 
	\put(-10,40){\framebox(24,20){3}\framebox(24,20){}\framebox(24,20){4}\framebox(24,20){}\framebox(24,20){2}} 
	\put(-10,20){\framebox(24,20){}\framebox(24,20){3}\framebox(24,20){}\framebox(24,20){4}\framebox(24,20){}} 
	\put(-10,0){\framebox(24,20){4}\framebox(24,20){}\framebox(24,20){3}\framebox(24,20){}\framebox(24,20){}} 
\end{picture}
\caption{\label{fig:beatSchmidt}The states of Eq.~(\ref{beatRj}), having $\sum_jR_j = 12$, exceeding the (one-way) bound of Theorem~\ref{DADBr}, $\sum_jR_j\le D_A\lfloor D_B/r\rfloor=10$ with $r=2$. See text for detailed two-way protocol.}
\end{figure}

Can this bound be exceeded when $r=1$ and the parties use two-way communication? As stated above, several results seem to suggest that the answer may well be negative \cite{Ghosh,GhoshPRL,Ghosh2,Horodecki2Sen2,Nathanson}. However, I now give a set of distinguishable states on $3\times3$ for which the sum of Schmidt ranks is $\sum R_j = 10 > 9 = D_AD_B$. The states are,
\begin{eqnarray}
	|\Psi_1\rangle & = & \frac{1}{\sqrt{2}}(|00\rangle_{AB} + |\Phi_02\rangle_{AB}) + (|0\rangle_{A} + |1\rangle_{A})|1\rangle_B, \nonumber \\
	|\Psi_2\rangle & = & \frac{1}{\sqrt{2}}(|00\rangle_{AB} + |\Phi_02\rangle_{AB}) - (|0\rangle_{A} + |2\rangle_{A})|1\rangle_B,\nonumber \\
	|\Psi_3\rangle & = & |10\rangle_{AB},~~~~~~~~~~~~~~|\Psi_5\rangle  =  |\Phi_12\rangle_{AB}, \nonumber \\
	|\Psi_4\rangle & = & |20\rangle_{AB},~~~~~~~~~~~~~~|\Psi_6\rangle  =  |\Phi_22\rangle_{AB},
\end{eqnarray}
with
\begin{eqnarray}
	|\Phi_0\rangle_A & = & \frac{1}{3}(|0\rangle_{A} + 2|1\rangle_{A} + 2|2\rangle_{A}), \nonumber \\
	|\Phi_1\rangle_A & = & \frac{1}{3}(2|0\rangle_{A} + |1\rangle_{A} - 2|2\rangle_{A}),\nonumber \\
	|\Phi_2\rangle_A & = & \frac{1}{3}(2|0\rangle_{A} - 2|1\rangle_{A} + |2\rangle_{A}),
\end{eqnarray}
forming an orthonormal basis. Bob starts with the following pair of measurement operators,
\begin{eqnarray}
	B_{1} & = & |0\rangle_{B}\langle 0| + \frac{1}{\sqrt{2}}|1\rangle_{B}\langle 1|, \nonumber \\
	B_{2} & = & |2\rangle_{B}\langle 2| + \frac{1}{\sqrt{2}}|1\rangle_{B}\langle 1|.
\end{eqnarray}
If Bob obtains $B_1$, $|\Psi_5\rangle$ and $|\Psi_6\rangle$ are excluded, $|\Psi_3\rangle$ and $|\Psi_4\rangle$ are unchanged, and (apart from unimportant normalization)
\begin{eqnarray}
	|\Psi_1\rangle & \rightarrow & |00\rangle_{AB} + (|0\rangle_{A} + |1\rangle_{A})|1\rangle_B, \nonumber \\
	|\Psi_2\rangle & \rightarrow & |00\rangle_{AB} - (|0\rangle_{A} + |2\rangle_{A})|1\rangle_B.
\end{eqnarray}
Alice follows with an orthogonal projective measurement onto the standard basis in her space. For each of her outcomes, only two states remain and are still orthogonal, and Bob can then distinguish which one they have.

When Bob obtains $B_2$, it turns out that the basic structure of the remaining states is exactly the same as for $B_1$, the only difference being that Alice must now measure in the basis of the orthogonal states, $|\Phi_k\rangle_A$. This is easily seen by recognizing that
\begin{eqnarray}
	 |0\rangle_{A} + |1\rangle_{A} & = |\Phi_0\rangle_A + |\Phi_1\rangle_A, \nonumber \\
	 |0\rangle_{A} + |2\rangle_{A} & = |\Phi_0\rangle_A + |\Phi_2\rangle_A ,
\end{eqnarray}
In fact, the structure of the original states was also the same, considered from the point of view of the $|\Phi_k\rangle_A$ basis as compared to Alice's standard basis. Hence, the parties can also distinguish the states with certainty for $B_2$, and the bound $\sum R_j \le D_AD_B$ has been exceeded. The generalization of this construction to higher dimensions will be discussed elsewhere \cite{tobeNote}.

For a $3\times 3$ system with $r\ge 2$, $N_{\max}=1$ so the Schmidt rank sum cannot exceed $3=D_A\lfloor D_B/r\rfloor$ even if LOCC-K2 is employed. Separable operations, on the other hand, allow this sum to be at least equal to $6$, as has already been demonstrated by the example of Eq.~(\ref{eqn:Yu}). It would be useful to have a (non-trivial) upper bound on the Schmidt rank sum for general LOCC and for SEP, but we are unable to provide one here.

\subsection{Preserving the original Schmidt ranks}

Given a set of states to be distinguished, perhaps the most difficult task, and the ideal outcome, would be to distinguish while preserving the original state intact. Failing this, it might nonetheless be possible to preserve the original Schmidt ranks. Here, we consider this problem and give, separately, a sufficient and then a necessary condition such a set must satisfy.

There is a sufficient condition which is almost trivially obvious: if all the reduced density operators are orthogonal on one side or the other then only one party need measure, and they can distinguish preserving $R_j$ using LOCC-P0. A less trivial sufficient condition is given below as Theorem \ref{ccsp}, in which I use the notion of a ``cascading sequence of partitions", defined as follows: starting with an arbitrary set of states and considering their reduced density operators $\{\rho_j^A\}$, partition these into disjoint subsets such that each $\rho_j^A$ is orthogonal to all those $\rho_k^A$ corresponding to states in different subsets; then partition each of these subsets into smaller subsets in the same way except by considering $\{\rho_j^B\}$; and so on back and forth for as many steps as is possible. We will call this partitioning ``complete" if each final subset consists of a single member.
\begin{CCSP} \label{ccsp} The set of states $\{|\Psi_j\rangle\}$ is perfectly distinguishable by LOCC while preserving $R_j$ provided these states can be completely partitioned by a cascading sequence, as defined above. Indeed, under these conditions, the state may be preserved unchanged by LOCC-P2 (LOCC-P1 if there are only two levels to the sequence, one for Alice and one for Bob).
\end{CCSP}
The proof of this theorem is quite simple. The parties need just perform orthogonal measurements projecting onto the union of the supports of the appropriate density operators in each subset: the first measurement is chosen to correspond to the first level of the partitioning sequence; the second measurement is chosen to correspond to the subsets descending directly from that subset identified by the outcome of the first measurement, etc. \hspace{\stretch{1}}$\blacksquare$

Given the reduced density operators of the states on both sides, the condition may be checked in a fairly straightforward way. For the first level of partition, start by placing $\rho_1^A$ in a first subset ${\cal S}_1$, and check to see if $\rho_2^A$ is orthogonal to it. If not, also include the latter in ${\cal S}_1$; and otherwise put it into ${\cal S}_2$. Now check $\rho_3^A$: if it is orthogonal to both $\rho_1^A$ and $\rho_2^A$, include it in a new subset; otherwise, include it with the one it is not orthogonal to --- if it is orthogonal to neither, then they must all be included in the same subset even if $\rho_1^A$ and $\rho_2^A$ are orthogonal to each other. Continue in this way until all states are partitioned into subsets. For subsequent levels of partition, start with each subset appearing on the previous level and partition that subset as described above for the first level. If the previous level was partitioned according to Alice's density operators, then for the next one use Bob's, and vice-versa. If this process can be continued until all subsets contain only a single state, then the states can be distinguished without being altered. If not, then one should check again, this time starting with Bob's side instead of Alice's. While not exactly simple, it is nonetheless a relatively straightforward procedure, which could be readily coded as an algorithm for numerical implementation.

\begin{figure}
\centering
\begin{picture}(144,160)
	\put(0,146){\makebox(24,20){$|0\rangle_B$}\makebox(24,20){$|1\rangle_B$}\makebox(24,20){$|2\rangle_B$}\makebox(24,20){$|3\rangle_B$}\makebox(24,20){$|4\rangle_B$}\makebox(24,20){$|5\rangle_B$}}
	\put(-24,126){\makebox(20,20){$|0\rangle_A$}}
	\put(-24,106){\makebox(20,20){$|1\rangle_A$}}
	\put(-24,86){\makebox(20,20){$|2\rangle_A$}}
	\put(-24,66){\makebox(20,20){$|3\rangle_A$}}
	{\thicklines\put(0,66){\framebox(144,80)}} 
	\put(0,126){\framebox(24,20){1}\framebox(24,20){}\framebox(24,20){2}\framebox(24,20){}\framebox(24,20){}\framebox(24,20){}} 
	\put(0,106){\framebox(24,20){}\framebox(24,20){1}\framebox(24,20){}\framebox(24,20){2}\framebox(24,20){5}\framebox(24,20){}}
	\put(0,86){\framebox(24,20){3,4}\framebox(24,20){}\framebox(24,20){3,4}\framebox(24,20){}\framebox(24,20){}\framebox(24,20){5}}
	\put(0,66){\framebox(24,20){}\framebox(24,20){3,4}\framebox(24,20){}\framebox(24,20){3,4}\framebox(24,20){}\framebox(24,20){}} 
	\put(-40,12){$\{1,2,3,4,5\}$}
	\put(8,18){$\Large{\nearrow}$}\put(12,12){Bob}\put(8,6){$\searrow$}
	\put(24,0){$\{\rho_5^B\}$}\put(24,24){$\{\rho_1^B,\rho_2^B,\rho_3^B,\rho_4^B\}$}
	\put(90,30){$\Large{\nearrow}$}\put(94,24){Alice}\put(90,18){$\searrow$}
	\put(104,44){$\{\rho_1^A,\rho_2^A\}$}\put(104,4){$\{\rho_3^A,\rho_4^A\}$}
	\put(138,50){$\Large{\nearrow}$}\put(144,44){Bob}\put(138,38){$\searrow$}
	\put(152,56){$\{\rho_1^B\}$}\put(152,32){$\{\rho_2^B\}$}
	\put(138,10){$\Large{\nearrow}$}\put(144,4){Bob}\put(138,-2){$\searrow$}
	\put(152,16){$\{\rho_3^B\}$}\put(152,-8){$\{\rho_4^B\}$}
\end{picture}
\vspace{.1in}
\caption{\label{fig:ccsp}Illustration of the procedure of cascading partitions for testing Theorem~\ref{ccsp}, described in the text (the states are given in Eq.~(\ref{eqn:ccsp})). Bob can separate out $\{\rho_5^B\}$ since it is orthogonal to all the others, after which Alice can divide the remaining four into $\{\rho_1^A,\rho_2^A\}$ and $\{\rho_3^A,\rho_4^A\}$. Then Bob can complete the partitioning.}
\vspace{.1in}
\end{figure}
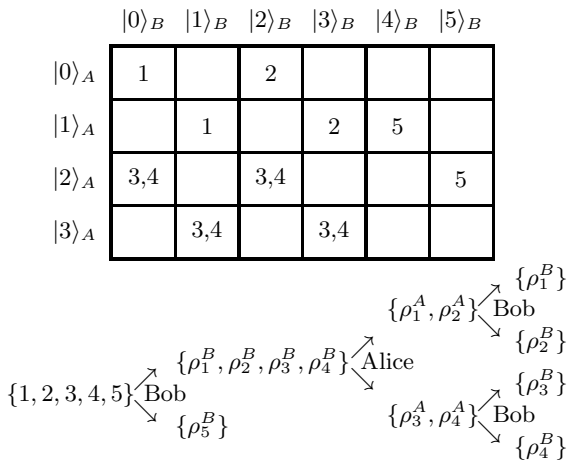
This procedure is illustrated by the following set of states, represented in Fig.~\ref{fig:ccsp}. We have
\begin{eqnarray}\label{eqn:ccsp}
|\Psi_1\rangle & = & |00\rangle_{AB}+|11\rangle_{AB},\nonumber\\
|\Psi_2\rangle & = & |02\rangle_{AB}+|13\rangle_{AB},\nonumber\\
|\Psi_3\rangle & = & |2\rangle_A(|0\rangle_B+|2\rangle_B)+|3\rangle_A(|1\rangle_B+|3\rangle_B),\nonumber\\
|\Psi_4\rangle & = & |2\rangle_A(|0\rangle_B-|2\rangle_B)+|3\rangle_A(|1\rangle_B-|3\rangle_B).\nonumber\\
|\Psi_5\rangle & = & |14\rangle_{AB}+|25\rangle_{AB}.
\end{eqnarray}
Notice first that Alice cannot start the procedure, since her density operators do not partition into two non-empty subsets such that all those in one subset are orthogonal to all those in the other. On the other hand, Bob can separate out $\{\rho_5^B\}$ since it is orthogonal to all the others, after which Alice can divide the remaining four into $\{\rho_1^A,\rho_2^A\}$ and $\{\rho_3^A,\rho_4^A\}$. Then Bob can complete the partitioning (his corresponding measurement will depend on the outcome of Alice's preceding one).

This condition is not a necessary one. It is not satisfied by the distinguishable set of product states (see Fig.~\ref{fig:notCCSP}),
\begin{figure}
\centering
\begin{picture}(60,80)
	\put(0,60){\makebox(24,20){$|0\rangle_B$}\makebox(24,20){$|1\rangle_B$}\makebox(24,20){$|2\rangle_B$}}
	\put(-24,40){\makebox(20,20){$|0\rangle_A$}}
	\put(-24,20){\makebox(20,20){$|1\rangle_A$}}
	\put(-24,0){\makebox(20,20){$|2\rangle_A$}}
	{\thicklines\put(0,0){\framebox(72,60)}} 
	\put(0,40){\framebox(24,20){1}\framebox(24,20){3}\framebox(24,20){3}} 
	\put(0,20){\framebox(24,20){1}\framebox(24,20){}\framebox(24,20){4}} 
	\put(0,0){\framebox(24,20){2}\framebox(24,20){2}\framebox(24,20){4}} 
\end{picture}
\caption{\label{fig:notCCSP}Demonstration that Theorem~\ref{ccsp} does not provide a necessary condition. These product states can be distinguished (so $R_j = 1$ is preserved in all cases), but they cannot be completely partitioned by a cascading sequence. In fact, they cannot be partitioned even once into two non-empty subsets where the reduced density operators in one subset are orthogonal to all those in the other.}
\vspace{.05in}
\end{figure}
\begin{eqnarray}
|\Psi_1\rangle & = & (|0\rangle_A+|1\rangle_A)|0\rangle_B,\nonumber\\
|\Psi_2\rangle & = & |2\rangle_A(|0\rangle_B+|1\rangle_B),\nonumber\\
|\Psi_3\rangle & = & |0\rangle_A(|1\rangle_B+|2\rangle_B),\nonumber\\
|\Psi_4\rangle & = & (|1\rangle_A+|2\rangle_A)|2\rangle_B.
\end{eqnarray}
Looking at Bob's reduced density operators, for example: the first is not orthogonal to the second, the second not to the third, etc. Since the same argument holds on Alice's side, these states cannot be partitioned even once into two non-empty subsets where the reduced density operators in one subset are orthogonal to all those in the other. Nonetheless, they can readily be distinguished by LOCC-P0 using projective measurements in the standard basis on both sides.

A necessary condition is given in the next theorem, stated in terms of a set of density operators defined as,
\begin{equation}
	\hat \rho_j = \rho_j^A\otimes\rho_j^B.
\end{equation}
\begin{keepRj} \label{KeepRj} If a set of states $\{|\Psi_j\rangle\}$ is perfectly distinguishable by LOCC while preserving $R_j$, then the density operators $\{\hat\rho_j\}$ form a mutually orthogonal set.
\end{keepRj}
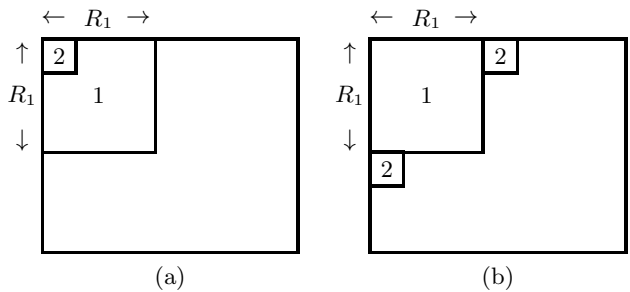
\begin{figure}
\centering
\begin{picture}(144,80)
	\put(-40,84){\makebox(42,8){$R_1$}}\put(-40,84){\makebox(72,8){$\rightarrow$}}\put(-40,84){\makebox(8,8){$\leftarrow$}}
	\put(84,84){\makebox(42,8){$R_1$}}\put(84,84){\makebox(72,8){$\rightarrow$}}\put(84,84){\makebox(8,8){$\leftarrow$}}
	\put(-52,38){\makebox(8,42){$R_1$}}\put(-52,54){\makebox(8,42){$\uparrow$}}\put(-52,38){\makebox(8,8){$\downarrow$}}
	\put(72,38){\makebox(8,42){$R_1$}}\put(72,54){\makebox(8,42){$\uparrow$}}\put(72,38){\makebox(8,8){$\downarrow$}}
	\put(-40,-14){\makebox(96,8){(a)}}
	\put(84,-14){\makebox(96,8){(b)}}
	{\thicklines\put(-40,0){\framebox(96,80)}} 
	{\thicklines\put(-40,38){\framebox(42,42){1}}}
	{\thicklines\put(84,0){\framebox(96,80)}} 
	{\thicklines\put(84,38){\framebox(42,42){1}}}
	{\thicklines\put(-40,68){\framebox(12,12){2}}}
	{\thicklines\put(84,25.25){\framebox(12,12){2}}}
	{\thicklines\put(127,68){\framebox(12,12){2}}}
\end{picture}
\vspace{.1in}
\caption{\label{fig:keepRj}Intuitive picture indicating how Theorem~\ref{KeepRj} can be proved. The parts of this figure correspond to the two ways it can happen that $\hat\rho_2\hat\rho_1 \ne 0$ (only selected components of $|\Psi_2\rangle$ are shown). In either case, any individual measurements the parties can perform that preserve $R_j$ leave the picture essentially unchanged, which means they have not distinguished. See Appendix~\ref{keepRj_proof} for a detailed proof.}
\end{figure}
The proof of this theorem is presented in Appendix~\ref{keepRj_proof}. The idea behind the proof can be seen from the following discussion. If $\hat\rho_2\hat\rho_1 \ne 0$, there are two possible ways this may come about, as indicated in parts (a) and (b) of Fig.~\ref{fig:keepRj}. The idea is that the two states are too closely intertwined in each case for them to be separated without significant distortion (that is, without a decrease in Schmidt rank). The first possibility is shown in part (a) of the figure, in which a component of $|\Psi_2\rangle$ lies within the $R_1\times R_1$ box representing the region of Hilbert space that is fully (at least according to the reduced density operators) occupied by $|\Psi_1\rangle$. As is shown for this case in Appendix~\ref{keepRj_proof}, neither party can ``remove" $|\Psi_2\rangle$ from the $|\Psi_1\rangle$ box by any complete LOCC measurement without reducing the Schmidt rank of one or the other of the states. Therefore, the picture shown in the figure persists throughout their protocol, no matter how many rounds of measurements they make. This means they can never eliminate $|\Psi_2\rangle$ while preserving $|\Psi_1\rangle$ and must fail to distinguish. The second case, illustrated in Fig.~\ref{fig:keepRj}(b), is argued in essentially the same way. Note that it may be possible for the parties to implement individual measurement operators that separate the states and preserve $R_j$, but it is not possible for them to do so for every outcome of a complete measurement.

Every set of orthogonal product states satisfies the conditions of this theorem, but it is well known not every such set can be distinguished by LOCC, demonstrating that the condition of the theorem is not a sufficient one. The best known example of such a set of product states was provided by Bennett and co-workers \cite{Bennett9} in their discussion of ``nonlocality without entanglement". Other proofs of this phenomenon, simplifying that of the original paper, have appeared in the literature \cite{WalgateHardy,Groisman}. However, it does not appear to this author that any of these proofs is particularly transparent or intuitive. In Appendix~\ref{app:Bennett9}, I supply such a proof, where it is shown in a very simple and direct way that the parties cannot perform any local operation other than a unitary without destroying the orthogonality of the states. Since a unitary operation cannot yield any information, nor can it eliminate even one of the states, the parties cannot distinguish this set of states.

The following example also demonstrates the condition of this theorem is not sufficient, this time with entangled states. The $\hat\rho_j$ are orthogonal, but they cannot be distinguished since there is no measurement either party can make that is less than full rank without reducing the Schmidt rank of at least one of the states. The states are
\begin{eqnarray}
	|\Psi_1\rangle & = & |01\rangle_{AB} + |12\rangle_{AB}, \nonumber \\
	|\Psi_2\rangle & = & |13\rangle_{AB} + |24\rangle_{AB},\nonumber \\
	|\Psi_3\rangle & = & |20\rangle_{AB} + |31\rangle_{AB}, \nonumber \\
	|\Psi_4\rangle & = & |32\rangle_{AB} + |43\rangle_{AB}.
\end{eqnarray}
For example, if $|\Psi_1\rangle$ is not eliminated for outcome $A_l$, then the support of $A_l$ must include a two-dimensional subspace that is not orthogonal to either $|0\rangle_A$ or $|1\rangle_A$. Then $|\Psi_2\rangle$ is not eliminated so the support of $A_l$ cannot be orthogonal to $|2\rangle_A$, etc. On the other hand, if $|\Psi_1\rangle$ is eliminated, then the kernel of $A_l$ must include $|0\rangle_A$ and $|1\rangle_A$, which means that $|\Psi_2\rangle$ must also be eliminated so the kernel of $A_l$ must include $|2\rangle_A$ as well, etc. Thus, since we may assume $A_l \ne 0$, the rank of $A_l$ must be $D_A$. It is true that the structure of these states is altered by this operation --- for example, $|k\rangle_A \rightarrow |a_k^l\rangle_A$ --- but while the $|a_k^l\rangle_A$ need not be orthogonal, they do need to be linearly independent. Then, an argument similar to the above will again show that subsequent measurements by the two parties must all be full rank. This means they can never eliminate even a single state, so this set cannot be distinguished without reducing at least one of them to a product state.

The condition of the theorem is not necessary for SEP. The states of Eq.~(\ref{eqn:Yu}) provide a counter-example, since as already shown they are distinguishable by SEP while preserving the original Schmidt ranks, but the density operators $\hat\rho_j$, corresponding to these states, are not mutually orthogonal. It is conceivable, on the other hand, that the condition of the theorem is sufficient for SEP, but we do not know if this is the case. Given the $\hat\rho_j$ are mutually orthogonal, one might try constructing a separable measurement starting with orthogonal projectors $A_j\otimes B_j$, one for each state $|\Psi_j\rangle$, such that the support of $A_j~(B_j)$ is equal to that of $\rho_j^A~(\rho_j^B)$. However, if the set of states is an unextendible product basis \cite{UPB_CMP,UPB_PRL} (each such set satisfies the conditions of the theorem), then the projector onto the remaining part of Hilbert space is proportional to a bound entangled state, meaning that no separable operation exists to complete this measurement. The starting point of this argument is a very special set of operations, so it does not constitute a proof the states are indistinguishable by SEP. In fact, it has been proven that every unextendible product basis in $3\times 3$ is distinguishable by SEP \cite{UPB_CMP}, so sufficiency for SEP remains an open question.

Since the rank of $\hat\rho_j$ is $R_j^2$ and we know from the previous theorem that deterministic distinguishing while always preserving $R_j$ requires the set of these density operators to be mutually orthogonal, we have
\begin{R2}\label{R2}
	If a set of states can be perfectly distinguished by LOCC while always preserving $R_j$, then
\begin{equation}
	\sum_{j=1}^NR_j^2\le D_AD_B.
\end{equation}
\end{R2}

Once again, the set of states in Eq.~(\ref{eqn:Yu}) provides an example showing that this corollary does not hold for SEP. For these states, $D_AD_B = 9$ whereas $\sum_jR_j^2 = 12$.

\section{Additional Theorems}
\label{resRank}

We now give two additional theorems, which relate Schmidt ranks of the states in the original set to be distinguished with those of the residual states. In particular, I consider how the largest Schmidt rank $r_j$ that can be preserved for state $|\Psi_j\rangle$ is constrained by the collection of original Schmidt ranks $\{R_{j^\prime}\}$ and the dimensions $D_A$, $D_B$ of the Hilbert spaces ${\cal H}_A$, ${\cal H}_B$. It will be convenient to write the original states as
\begin{eqnarray}
	\label{ensemble}
	|\Psi_j \rangle & = & \sum_{m,n=1}^{D}({M}_j)_{nm} |m\rangle_A |n\rangle_B,
\end{eqnarray}
with ${M}_j$ a matrix of rank $R_j$. Then for a given measurement outcome, $A \otimes B$, the parties will be left with
\begin{equation}
	\label{twoway}
	A \otimes B|\Psi_j\rangle = \sum_{m,n=1}^{D}({B}{M}_j{A^T})_{nm} |m\rangle_A |n\rangle_B,
\end{equation}
where $A^T$ is the transpose of the matrix $A$. In the following, we will use two facts:
\begin{enumerate}
	\item the Schmidt rank, $r_j$, of the residual state is given by the rank of the matrix $r({B}{M}_j{A^T})$;
and 
	\item if $|\Psi_j\rangle$ is identified deterministically (or unambiguously \cite{Sun}) by outcome $A \otimes B$, then ${B}{M}_k{A^T}=0~\forall_{k\ne j}$.
\end{enumerate}
We will also find useful in this section two inequalities on matrix ranks \cite{HornJohnson}, which say that for $m\times l$ matrix $X$ and $l\times n$ matrix Y, the rank $r(XY)$ of their product is bounded as
\begin{equation}
	\label{matrixRankbound}
	\min[r(X),r(Y)]\ge r(XY) \ge r(X) + r(Y) - l.
\end{equation}
The first theorem we will consider concerns general, two-way protocols and applies in both the deterministic and unambiguous cases. 
\begin{Dby2} \label{DBy2} Given the task of deterministically or unambiguously distinguishing a set of bipartite states, $\{|\Psi_j\rangle\}$ having Schmidt ranks $\{R_j\}$, then for every separable outcome $A_m\otimes B_m$ distinguishing $|\Psi_j\rangle$ and preserving $r_j^m$,
\begin{equation}
	2r_j^m  + \max_{k\ne j}(R_k) \le D_A + D_B.
\end{equation}
\end{Dby2}
\noindent Proof: For either the deterministic or unambiguous case, we have that $r_j^m = r({B}_m{M}_j{A}_m^T)$ implying $r({B}_m) \ge r_j^m$ and $r({A}_m) \ge r_j^m$, and $r({B}_m{M}_k{A}_m^T)=0~\forall_{k\ne j}$. From the latter expression with Eq.~(\ref{matrixRankbound}), we have
\begin{eqnarray}
	0 & \ge & r({B}_m) + r({M}_k{A}_m^T) - D_B \nonumber \\
	& \ge & r({B}_m) + r({A}_m) + r({M}_k) - D_A - D_B\nonumber \\
	& \ge & 2r_j^m + R_k - D_A - D_B.
\end{eqnarray}
and the theorem easily follows.\hspace{\stretch{1}}$\blacksquare$

Note how this expression explicitly shows the tradeoff between the original and final Schmidt ranks, in relationship to the Hilbert space dimensions. The following corollary offers one example of how this result can be useful.
\begin{Dby2C} \label{DBy2C} If any of the original states, say the first, has Schmidt rank $R_1 = D_A$ ($D_A \le D_B$), then one cannot preserve Schmidt rank exceeding $D_B/2$ for any single SEP outcome identifying $|\Psi_j\rangle$ with certainty when $j\ne 1$. If any two states start out with Schmidt ranks equal to $D_A$, then no outcome can preserve greater than $D_B/2$.
\end{Dby2C}
When $D_B/2 < D_A$, these statements are non-trivial and are a consequence of the extent to which the rank-$D_A$ states are spread through the space, so they cannot be annihilated by measurement operators of rank exceeding $D_B/2$. In general, as the largest Schmidt rank $R_{\max}$ decreases, less of the space is occupied by the corresponding state, which can then be annihilated by higher-rank operators, allowing larger Schmidt rank to be preserved for other states. From another point of view, the amount of information required to distinguish decreases along with $R_{\max}$, so the parties may use less refined measurements allowing $r_j^m$ to be greater.

According to the proof of this theorem, if any single state $|\Psi_\kappa\rangle$ is excluded by the outcome $m$, even if there are non-zero probabilities for identifying several other states, the bound in the theorem still holds with $\max_{k\ne j}(R_k)$ replaced by $R_\kappa$. Then the result becomes applicable to protocols that allow for errors in identifying the state.

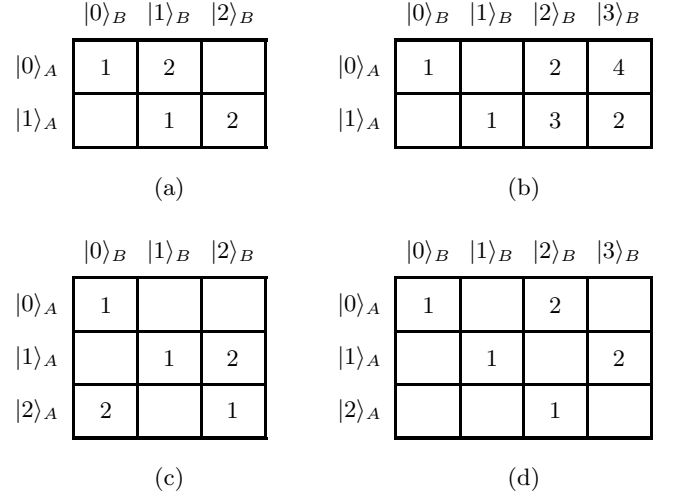
\begin{figure}
\centering
\begin{picture}(144,166)
	\put(-26,160){\makebox(24,20){$|0\rangle_B$}\makebox(24,20){$|1\rangle_B$}\makebox(24,20){$|2\rangle_B$}}
	\put(-50,120){\makebox(20,20){$|1\rangle_A$}}
	\put(-50,140){\makebox(20,20){$|0\rangle_A$}}
	{\thicklines\put(-26,120){\framebox(72,40)}} 
	\put(-26,140){\framebox(24,20){1}\framebox(24,20){2}\framebox(24,20){}} 
	\put(-26,120){\framebox(24,20){}\framebox(24,20){1}\framebox(24,20){2}}
	\put(-38,100){\makebox(96,8){(a)}}
	\put(96,160){\makebox(24,20){$|0\rangle_B$}\makebox(24,20){$|1\rangle_B$}\makebox(24,20){$|2\rangle_B$}\makebox(24,20){$|3\rangle_B$}}
	\put(72,120){\makebox(20,20){$|1\rangle_A$}}
	\put(72,140){\makebox(20,20){$|0\rangle_A$}}
	{\thicklines\put(96,120){\framebox(96,40)}} 
	\put(96,140){\framebox(24,20){1}\framebox(24,20){}\framebox(24,20){2}\framebox(24,20){4}} 
	\put(96,120){\framebox(24,20){}\framebox(24,20){1}\framebox(24,20){3}\framebox(24,20){2}} 
	\put(96,100){\makebox(96,8){(b)}}
	\put(-26,70){\makebox(24,20){$|0\rangle_B$}\makebox(24,20){$|1\rangle_B$}\makebox(24,20){$|2\rangle_B$}}
	\put(-50,10){\makebox(20,20){$|2\rangle_A$}}
	\put(-50,30){\makebox(20,20){$|1\rangle_A$}}
	\put(-50,50){\makebox(20,20){$|0\rangle_A$}}
	{\thicklines\put(-26,10){\framebox(72,60)}} 
	\put(-26,50){\framebox(24,20){1}\framebox(24,20){}\framebox(24,20){}} 
	\put(-26,30){\framebox(24,20){}\framebox(24,20){1}\framebox(24,20){2}}
	\put(-26,10){\framebox(24,20){2}\framebox(24,20){}\framebox(24,20){1}}
	\put(-38,-10){\makebox(96,8){(c)}}
	\put(96,70){\makebox(24,20){$|0\rangle_B$}\makebox(24,20){$|1\rangle_B$}\makebox(24,20){$|2\rangle_B$}\makebox(24,20){$|3\rangle_B$}}
	\put(72,10){\makebox(20,20){$|2\rangle_A$}}
	\put(72,30){\makebox(20,20){$|1\rangle_A$}}
	\put(72,50){\makebox(20,20){$|0\rangle_A$}}
	{\thicklines\put(96,10){\framebox(96,60)}} 
	\put(96,50){\framebox(24,20){1}\framebox(24,20){}\framebox(24,20){2}\framebox(24,20){}} 
	\put(96,30){\framebox(24,20){}\framebox(24,20){1}\framebox(24,20){}\framebox(24,20){2}} 
	\put(96,10){\framebox(24,20){}\framebox(24,20){}\framebox(24,20){1}\framebox(24,20){}} 
	\put(96,-10){\makebox(96,8){(d)}}
\end{picture}
\vspace{.1in}
\caption{\label{fig:rRbound}Illustration of Theorem~\ref{DBy2}. (a) For $D_A = 2$ and $D_B = 3$, it is not possible to preserve entanglement while distinguishing if more than one state is initially entangled. (b) Increasing $D_B$ allows higher-rank states to be fully separated from each other, and then preserving entanglement becomes possible. As seen in the right half of this diagram, however, the presence of additional states can alter this conclusion. (c) For a $3\times3$ system with $R_1 = 3$, $|\Psi_2\rangle$ cannot remain entangled after being identified by the measurements. (d) Again, increasing the size of ${\cal H}_B$ allows more space for the states and even with $R_1=3$ it is possible to preserve $r_2=2$.}
\end{figure}
We see from this theorem that when $D_A=D_B=2$, it is not possible to distinguish by SEP while preserving entanglement, even for a single outcome (this conclusion holds for all protocols that exclude at least one state for every final outcome, so is not restricted to deterministic, or even unambiguous, distinguishing). Other simple examples giving an intuitive picture for this theorem are presented in Fig.~\ref{fig:rRbound}. For the case $D_A = 2$ and $D_B = 3$ ($D_B/2<D_A$), preserving entanglement while distinguishing requires that no more than one state is initially entangled.  In part (a) of the figure, we see that with both states rank-$2$, they are necessarily too intertwined for entanglement to be preserved. Increasing $D_B$ allows higher-rank states to be fully separated from each other, and preserving entanglement becomes possible (Fig.~\ref{fig:rRbound}(b)). With a $3\times3$ system no more than one of the states can have $R_j=3$, and if one does have this rank, the other states cannot remain entangled after being identified by the measurements. This is seen in Fig.~\ref{fig:rRbound}(c) where $|\Psi_1\rangle$ can be distinguished in the $|0\rangle_A|0\rangle_B+|1\rangle_A|1\rangle_B$ corner of the box, preserving rank-$2$, but no other outcomes can distinguish while preserving entanglement. Again, increasing the size of ${\cal H}_B$ allows more space for the states and even with $R_1=3$ it is possible to preserve $r_2=2$ (Fig.~\ref{fig:rRbound}(d)). Notice that in Fig.~\ref{fig:rRbound}(b) and (d), Schmidt rank of $D_B/2$ can be preserved. In Fig.~\ref{fig:rRbound}(c) we see that $D_B/2$ can be exceeded for $r_1$, but this is in line with the theorem, since the rank of the other state is $R_2<D_A$. It is easily seen in the latter case that any attempt to increase $R_2$ to $D_A$ would destroy the ability to preserve $r_1>D_B/2$.

When $D_A$ and $D_B$ do not differ by too much, the theorem gives a non-trivial bound, but since Schmidt ranks cannot exceed the smaller dimension, this is no longer the case for $D_B/2\ge D_A$ (or with $A$ and $B$ reversed). In addition, as is illustrated in the right half of Fig.~\ref{fig:rRbound}(b), the presence of additional states can alter conclusions about the amount of entanglement it is possible to preserve. This demonstrates that the theorem gives only a necessary, and not a sufficient, condition for preservation of entanglement. 

We now consider a restriction to one-way classical communication. The next theorem again shows there is a tradeoff between the starting and residual Schmidt ranks, though here the tradeoff involves both the number of states, $N$, and the average Schmidt rank, $\overline R=\sum_j R_j/N$ \cite{Th1stronger}.

\begin{DbyN}\label{DbyN} With Alice going first in a one-way LOCC protocol, if for any one of Alice's outcomes ($A_m$) Bob is able to deterministically distinguish the remaining states, then
\begin{eqnarray}
	\label{eq:DbyN}
	r_j^m + \overline R \le D_A + D_B/N,
\end{eqnarray}
where $r_j^m$ refers here to the Schmidt rank of $|\Psi_j\rangle$ following Alice's outcome, and both before and after Bob measures (see below).
\end{DbyN}
\noindent Proof: In order for Bob to be able to distinguish with certainty after Alice obtains outcome $m$, the reduced density operators of the various possible states remaining must be mutually orthogonal (implying that they can preserve $r_j^m$). The rank of each of these reduced density operators is $r({M}_j{A}_m^T)$ (Bob has yet to do anything so I have set ${B} = I_B$), and their orthogonality implies that the sum of these ranks cannot exceed $D_B$. Then, again using Eq.~(\ref{matrixRankbound}), we have
\begin{eqnarray}
	D_B & \ge & \sum_{j=1}^N r({M}_j{A}_m^T) \nonumber \\
	& \ge & \sum_{j=1}^N (R_j + r({A}_m) - D_A) \nonumber \\
	& = & \sum_{j=1}^N R_j + N(r({A}_m) - D_A).
\end{eqnarray}
With $r_j^m \le r({A}_m)$, the theorem follows immediately. \hspace{\stretch{1}}$\blacksquare$

The following upper bound on the number of states will follow as a direct consequence of this theorem:
\begin{Dbyr}\label{Dbyr}With Alice going first followed by Bob deterministically  distinguishing,
\begin{eqnarray}
	N \le \frac{D_B}{r_{\max} + \overline R - D_A},
\end{eqnarray}
with $r_{\max}$ the largest value of $r_j^m$.
\end{Dbyr}
Of course, this bound should only be applied if $r_{\max} + \overline R > D_A$; otherwise Eq.~(\ref{eq:DbyN}) is trivially satisfied without regard to the value of $N$.

If $D_A\le D_B$ and all the states have their maximum rank of $R_j=D_A$, we see that the parties cannot preserve Schmidt rank greater than $D_B/N$, or alternatively the number of states cannot exceed $D_B/r_{\max}$. If they are not concerned with preserving entanglement, then setting $r_{\max} = 1$ shows that if a set of Schmidt rank $D_A$ states can be distinguished perfectly by one-way LOCC, it cannot have more than $D_B$ members. This generalizes (at least when restricted to one-way communication) the results of \cite{GhoshPRL,Ghosh2,Nathanson} that no more than $D$ maximally entangled states on $D\times D$ can be perfectly distinguished. Fig.~\ref{fig:DbyN} makes clear that $D_B$ rank-$D_A$ states can be distinguished. These states fill the space, in the sense that the sum of Schmidt ranks is equal to the Hilbert space dimension. This diagram makes it seem almost intuitively obvious that adding another rank-$D_A$ state would make it impossible to distinguish (a conclusion which is correct, though when dealing with quantum systems, we should always be careful about trusting such intuitions).

In Appendix~\ref{2vs1}, it is shown that these bounds for one-way CC (Theorem~\ref{DbyN} and Corollary~\ref{Dbyr}) can be exceeded if two-way CC is allowed. Included in this appendix are examples where, depending on the outcome of Bob's measurement, Alice risks (1) by measuring, the destruction of entanglement that would otherwise be preserved; as opposed to (2) being unable to distinguish the states if she does not measure. Thus, we have interesting and non-trivial cases where the main purpose of the classical communication is simply to determine whether the next party should proceed with any measurment at all. 

In Appendix~\ref{cn}, two additional theorems are given, related to protocols of type LOCC-K0, where the parties are not allowed to communicate until after they complete their measurements. These theorems address the question of always distinguishing with a set of $N$ rank-$D$ states on $D\times D$, in which case Theorem~\ref{DbyN} tells us that the maximum possible residual Schmidt rank is $\lfloor D/N\rfloor$. One of these theorem shows that when this maximum rank is an integer, and the parties can preserve this rank for any single outcome, then they do so for all their outcomes using LOCC-P0. The other theorem deals with the case of non-integer $\lfloor D/N\rfloor$.

\begin{figure}
\centering
\begin{picture}(130,72)
	\put(-36,60){\makebox(36,20){$|0\rangle_B$}\makebox(36,20){$|1\rangle_B$}\makebox(20,20){$|2\rangle_B$}}
	\put(76,60){\makebox(36,20){$|D_B-3\rangle$}}
	\put(112,60){\makebox(36,20){$|D_B-2\rangle$}}
	\put(148,60){\makebox(36,20){$|D_B-1\rangle$}}
	\put(-60,0){\makebox(20,20){$|2\rangle_A$}}
	\put(-60,20){\makebox(20,20){$|1\rangle_A$}}
	\put(-60,40){\makebox(20,20){$|0\rangle_A$}}
	{\thicklines\put(-36,0){\framebox(92,60)}} 
	{\thicklines\put(76,0){\framebox(108,60)}} 
	\put(-36,40){\framebox(36,20){1}\framebox(36,20){2}\framebox(20,20){3}\makebox(20,20){$\cdots$}\framebox(36,20){$D_B-2$}\framebox(36,20){$D_B-1$}\framebox(36,20){$D_B$}} 
	\put(-36,20){\framebox(36,20){$D_B$}\framebox(36,20){1}\framebox(20,20){2}\makebox(20,20){$\cdots$}\framebox(36,20){$D_B-3$}\framebox(36,20){$D_B-2$}\framebox(36,20){$D_B-1$}} 
	\put(-36,0){\framebox(36,20){$D_B-1$}\framebox(36,20){$D_B$}\framebox(20,20){1}\makebox(20,20){$\cdots$}\framebox(36,20){$D_B-4$}\framebox(36,20){$D_B-3$}\framebox(36,20){$D_B-2$}} 
\end{picture}
\caption{\label{fig:DbyN}$D_B$ rank-$D_A$ states that are distinguishable by one-way LOCC.}
\end{figure}

\section{Discussion}\label{disc}
\subsection{Relationship to non-collective entanglement purification}
\label{entPure}
When entanglement is shared between two parties under realistic circumstances, it is very difficult to completely eliminate the effects of noise, which may enter in the creation of the entangled state or when it is shared between the parties through a quantum channel. As a result, the parties commonly share a mixed state rather than a pure one. Pure state entanglement is, however, necessary for many implementations of quantum information processing so it is important to understand when the parties will be able to purify their shared state.

Non-collective entanglement purification \cite{HorodeckiX3,Kent,MassarLinden} is the process of obtaining a pure entangled state from a single copy of a mixed state. The question we are considering in this paper is directly related to this process: Alice and Bob are given a state $|\Psi_j\rangle$ drawn from a set of $N$ mutually orthogonal, bipartite states with some \textit{a priori} probabilities $p_j$, but are not told which state was chosen. They may then describe their system by the mixed state,
\begin{equation}
	\rho = \sum_{j=1}^N p_j |\Psi_j\rangle\langle\Psi_j|.
\end{equation}
Together, with some probability, they perform an operation $\Gamma$, obtaining the new state
\begin{equation}
	\rho^\prime = \sum_{j=1}^N q_j |\Phi_j\rangle\langle\Phi_j|,
\end{equation}
where $|\Phi_j\rangle = \Gamma |\Psi_j\rangle/\sqrt{\langle\Psi_j|\Gamma^\dagger\Gamma|\Psi_j\rangle}$ and $q_j = p_j\langle\Psi_j|\Gamma^\dagger\Gamma|\Psi_j\rangle/\sum_{j=1}^N p_j\langle\Psi_j|\Gamma^\dagger\Gamma|\Psi_j\rangle$.

We want to know if and when $\rho^\prime$ is a pure entangled state. It will certainly be pure if $q_j = 0$ for all $j$ except one; that is, if $\Gamma|\Psi_j\rangle \sim \delta_{jJ}$, for some fixed $J$. Whether or not it is entangled will depend on the relationship between the operator $\Gamma$ and the state $|\Psi_J\rangle$. If it is entangled, then the parties have identified the original state as $|\Psi_J\rangle$ while preserving entanglement, which is the subject of the work described in this paper.

Can $\rho^\prime$ be pure and entangled when more than one of the $q_j$ are nonzero? The answer is yes if and only if the nonzero $\Gamma |\Psi_j\rangle$ are all the same, up to normalization and phase. In other words, entanglement purification is possible without distinguishing amongst the eigenstates of $\rho$, but only if there exists a product operator $\Gamma$ satisfying the above-stated condition. This is equivalent to the statement that there must exist a product projector such that all the original states not annihilated by it were ``equivalent" on the support of that projector. When such a projector does not exist, then the only possibility for non-collective entanglement purification is by the methods discussed in this paper.

Previous discussions of entanglement purification \cite{HorodeckiX3,Kent,MassarLinden} have focused on the case where the final state is uniformly entangled (all Schmidt coefficients equal to each other), corresponding to a maximally entangled state on a smaller space. The results presented in this paper are concerned instead only with the Schmidt rank of the residual state. However, given any pure entangled state, one can with nonzero probability obtain a uniformly entangled state by local operations on the separate parts \cite{BennettConcentrate,OurAtemp}. Therefore the questions addressed in this paper, concerning use of LOCC to distinguish a set of states and preserve entanglement, are directly related to previous discussions of non-collective entanglement purification.

\subsection{Multipartite systems} 

Any multipartite system may be viewed as bipartite by choosing a division of the parties into two groups. We may, for example, let Alice be in one group by herself and the remaining parties in the other. A question of interest is whether there is a way to extend our various arguments to also apply when the many parties are viewed separately. In order for this to be possible, we must first find a suitable generalization of Schmidt rank. Possible generalizations have been proposed \cite{Eisert,ChenLi}, but here I will only consider a rather simple one: the (generalized) Schmidt rank, $R_j$, of a multipartite quantum state is the smallest rank of the completely reduced density operators $\rho_j^A,~\rho_j^B,~\rho_j^C,~\cdots$.

The proof of Theorem~\ref{DADBn} involves each party locally and sequentially dividing the composite Hilbert space ${\cal H}_A \otimes {\cal H}_B$ into orthogonal subspaces. This division follows directly from the fact that their measurement operators include ones with non-empty kernel, which is orthogonal to that operator's support. Such divisions are not restricted to bipartite systems, and it is also true for the multipartite case that these divisions can be continued until all subspaces are no larger than $r\times r\times r\times \cdots$. If they must preserve (generalized) Schmidt rank of at least $r$ for each of their outcomes then after optimally choosing their operations, no part of any state can reside in any of the subspaces smaller than this, one state may be placed entirely within each of those that are $r\times r\times r\times \cdots$, and we see that
\begin{equation}\label{multiN}
	N_{max} = \lfloor D_A/r\rfloor\lfloor D_B/r\rfloor\lfloor D_C/r\rfloor\cdots
\end{equation}
for any multipartite system, when the parties must preserve at least $r$.

The proof of Theorem \ref{ccsp} for the multipartite case is also essentially the same as for bipartite systems, and the statement of the theorem applies without alteration. That is, if the multipartite states can be completely partitioned by a cascading sequence, then they can be perfectly distinguished while preserving the original states intact. The partitioning of the states into subsets again points to a protocol the parties may use, involving orthogonal projections onto combined supports of reduced density operators in the appropriate subsets.

\section{Summary} \label{sum}
In summary, I have introduced the question of preserving entanglement in the course of locally distinguishing an unknown state drawn from a set of orthogonal states. Several results on this topic have been proved. Theorem~\ref{DADBn}  (generalized by Eq.~(\ref{multiN})) gives the achievable maximum number of states on a multipartite system when a (generalized) Schmidt rank of $r$ must always be preserved. Theorem~\ref{DADBr} showed that for bipartite systems and one-way classical communication from Alice to Bob, the sum of Schmidt ranks of the states cannot exceed $D_A\lfloor D_B/r\rfloor$, when once again the parties must always preserve Schmidt rank of $r$. The next two theorems considered the possibility of preserving the original Schmidt ranks of the states in the set. Theorem~\ref{ccsp} applies to multipartite systems, and gives a sufficient condition that the states can be preserved unchanged. Theorem~\ref{KeepRj} then gives a necessary condition for preserving the original Schmidt ranks, that the set of density operators $\rho_j^A\otimes\rho_j^B$ must be mutually orthogonal. Following these results, I then proved two theorems that show explicitly a necessary relationship between the initial and final Schmidt ranks, given the parties must always distinguish the state.

In each case, we discussed how altering restrictions on the resources (types of operations and amount of classical communication) available to the parties may change these results. Various examples were provided illustrating this question, including explicit demonstrations of the superiority of two-way classical communication over protocols where the communication is restricted to be in only one direction. In particular, it was shown that the sum of Schmidt ranks can exceed the dimension of ${\cal H}_A \otimes {\cal H}_B$.
It may be recalled that Lo and Popescu \cite{LoPopescu} have shown that when locally manipulating pure states, anything that can be done using two-way communication can just as well be done with one-way communication alone. For the task of distinguishing a set of states and preserving entanglement, we have seen that one-way communication may not be sufficient. The reason is that we are effectively manipulating a mixed state. This latter point was discussed in the previous section, where I argued that the question of distinguishing while preserving entanglement is closely related to that of purifying entanglement from a mixed state.

\vspace{.15in}\noindent \textit{Acknowledgments} --- I would like to thank Yuqing Sun, Shengjun Wu, and especially Bob Griffiths for numerous enlightening discussions, both specifically on the topic of this paper, as well as on quantum information in general. I am also very grateful to Li Yu for assistance in counting the number of $r\times r$ subspaces for the proof of Theorem~\ref{DADBn}, and for sharing the extremely useful set of states and corresponding separable POVM given in Eqs.~(\ref{eqn:Yu}) and (\ref{eqn:POVM}). This work was partially supported by the National Science Foundation through Grant PHY-0456951.
\vspace{.15in}

\appendix

\section{Proofs of the theorems}
\label{proofs}

\subsection{Maximum number of states}
\label{nmax_proof}
$\\$
\noindent {\bf Theorem \ref{DADBn}} \textit{Suppose the parties share a $D_A \times D_B$ system and using LOCC-K2 are able to distinguish with certainty amongst a set of $N$ states while preserving Schmidt rank of at least $r$ for every outcome. Then,} 
\begin{equation}
	N \le \lfloor D_A/r\rfloor \lfloor D_B/r\rfloor \equiv N_{\max},
\end{equation}
\textit{where $\lfloor x\rfloor$ is the largest integer not greater than $x$, and this upper bound is achievable by LOCC-P0.}

\noindent Proof: At some point in the protocol one of the parties must implement a measurement operator that is less than full rank. The reason for this is that in order to eliminate any single state in the given set, say $|\Psi_J\rangle$, it must be that $(A \otimes B)|\Psi_J\rangle = 0$, for some $A$ and $B$. This means either $A$ or $B$ must be singular, so has non-trivial kernel. If it is Alice who first implements a singular operator, then that operator divides Alice's Hilbert space ${\cal H}_A$ into two orthogonal parts, its support and its kernel, of dimensions $D_{A1}$ and $D_A - D_{A1}$, respectively. If there are $N_1$ states that are not excluded, then these states must be distinguishable within the remaining $D_{A1} \times D_B$ dimensional space. Furthermore, the $N - N_1$ states that were excluded lie, from the outset, entirely in the other $(D_A - D_{A1}) \times D_B$ dimensional space, so \textit{at least this many} states must be distinguishable in that space. If we define a function $f(D_A,D_B)$ to be the maximum number of states perfectly distinguishable while preserving Schmidt rank at least $r$ in $D_A \times D_B$, then $N_1 \le f(D_{A1},D_B)$ and $N - N_1 \le f(D_A - D_{A1},D_B)$. Clearly, 
\begin{equation}
	N = N_1 + (N - N_1) \le f(D_{A1},D_B) + f(D_A - D_{A1},D_B).
\end{equation}
\noindent The maximum number of states in the original set is then bounded above as
\begin{equation}
	f(D_A,D_B) \le \max_{A_k} [f(D_{A1},D_B) + f(D_A - D_{A1},D_B)],
\end{equation}
\noindent and the maximum is taken over all choices of Alice's operator ${A_k}$; in other words, over all ways that she can divide her space into two orthogonal pieces.

We now look for upper bounds on $f(D_{A1},D_B)$ and $f(D_A - D_{A1},D_B)$ by considering measurements by Bob (it could just as well be Alice again) for each of the cases. Note that these measurements should be considered as completely unrelated protocols; each step in this argument involves a ``first" measurement (corresponding to a singular Kraus operator) in a new protocol aimed at distinguishing a smaller number of states on a smaller space. At the second step, we obtain
\begin{eqnarray}
	f(D_A,D_B) & \le & \max_{B_m}\{\max_{A_k} [f(D_{A1},D_{B1}) \nonumber \\ 
		& + & f(D_{A1},D_B - D_{B1})]  \nonumber \\
		& + & \max_{A_k} [f(D_A - D_{A1},D_{B1}^\prime) \nonumber \\
		& + & f(D_A - D_{A1},D_B - D_{B1}^\prime)]\},
\end{eqnarray}
\noindent and after many steps,
\begin{eqnarray}
	\label{maxr}
	f(D_A,D_B) \le \max[\sum_{l=1}^n f(D_{Al},D_{Bl})],
\end{eqnarray}
\noindent with the maximum now taken over operators that sequentially (and by local measurements) divide the original space into $n$ subspaces.

Since each division represents a successful outcome, one of the two subspaces at each step must be at least $r \times r$. If any subspace is larger than this, it can be divided by a subsequent measurement, so it is valid to continue the process until all subspaces are smaller than or equal to $r \times r$. Then the maximum in the above equation means choosing the best way to divide the space into such subspaces. Note that $f(D_{Al},D_{Bl}) = 1$ if both $D_{Al}$ and $D_{Bl}$ are equal to $r$ (Corollary~\ref{DBy2C}), and vanishes if either is less than $r$. Hence, the right-hand side of Eq.~(\ref{maxr}) is equal to the maximum number of orthogonal $r \times r$ subspaces in the original space. We can see $N_{\max}$ is an upper bound on this number by assuming otherwise and showing this leads to a contradiction. This assumption may be written $N = N_1 + (N-N_1)> \lfloor D_A/r\rfloor \lfloor D_B/r\rfloor$, leading to
\begin{eqnarray}
	N_1 + (N-N_1) > (\lfloor D_{A1}/r\rfloor + \lfloor (D_A-D_{A1})/r\rfloor)\lfloor D_B/r\rfloor,\nonumber\\
\end{eqnarray}
with $N$ and $N_1$ defined above. This implies either $N_1>\lfloor D_{A1}/r\rfloor\lfloor D_B/r\rfloor$ or $N-N_1>\lfloor (D_A-D_{A1})/r\rfloor\lfloor D_B/r\rfloor$. Following along the argument presented in the preceding part of this proof, one eventually arrives at a division for which one subspace is $r\times r$ and the other is no larger than this. We then have that there are either at least two states in an $r\times r$ subspace ($N_1>\lfloor r/r\rfloor\lfloor r/r\rfloor=1$), or at least one in a subspace smaller than this. This is a contradiction, giving us the stated upper bound. Certainly, there is no problem fitting $N_{\max}$~ $r\times r$ subspaces into the space, so the bound can be achieved, completing the proof.\hspace{\stretch{1}}$\blacksquare$

\subsection{Schmidt rank sum for one-way protocols}
\label{sec:Schmidt_proof}
\noindent {\bf Theorem \ref{DADBr}} \textit{If Alice goes first using LOCC-K1 and the parties are always able to distinguish and preserve Schmidt rank at least $r$, then $\sum_{j} R_j \le D_A \lfloor D_B/r\rfloor$.}

\noindent Proof: In order for Bob to be able to distinguish with certainty following Alice's measurement, the reduced density operators $\widetilde\rho_j^B$ of the various possible states remaining after Alice's outcome must be mutually orthogonal. If they must preserve Schmidt rank at least $r$ for each outcome, then each of these density operators must have rank at least $r$. These two requirements imply that for each of Alice's outcomes, $A_k$, no more than $\lfloor D_B/r\rfloor$ of the $|\Psi_j\rangle$ can have nonzero probability, or non-vanishing $(A_k^\dagger A_k \otimes I_B) |\Psi_j\rangle$. Let the eigenstate corresponding to nonzero eigenvalue $\lambda_m^k$ of $A_k^\dagger A_k$ be $|a_m^k\rangle$. Then,
\begin{eqnarray}
	(A_k^\dagger A_k \otimes I_B) |\Psi_j\rangle = \sum_m\lambda_m^k (|a_m^k\rangle\langle a_m^k|\otimes I_B)|\Psi_j\rangle,
\end{eqnarray}
\noindent which vanishes if and only if each term in the sum vanishes. Thus, no more than $\lfloor D_B/r\rfloor$ of the $|\Psi_j\rangle$ can satisfy $(|a_m^k\rangle\langle a_m^k| \otimes I_B) |\Psi_j\rangle \ne 0$ for any single eigenstate of Alice's POVM elements.

From the collection of eigenstates for all these POVM elements, choose a (generally non-orthogonal) basis, denoted by $\{|a_m\rangle\}$. Expanding the $|\Psi_j\rangle$ in the dual basis $\{|\overline a_m\rangle\}$, where $\langle a_m|\overline a_{m^\prime}\rangle=\delta_{mm^\prime}$,
\begin{eqnarray}
	|\Psi_j\rangle=\sum_{m=1}^{D_A} \mu_{m}^j |\overline a_{m}\rangle |b_{m}^j\rangle,
\end{eqnarray}
\noindent we see from the arguments of the previous paragraph that no more than $\lfloor D_B/r\rfloor$ of the $\mu_m^j$ can be non-zero, for any fixed $m$.

Now consider the Schmidt ranks,
\begin{eqnarray}
	R_j&=&R(|\Psi_j\rangle)=R(\sum_{m=1}^{D_A} \mu_{m}^j |\overline a_{m}\rangle |b_{m}^j\rangle) \nonumber \\
		&\le&\sum_{m=1}^{D_A}R(\mu_{m}^j |\overline a_{m}\rangle |b_{m}^j\rangle).
\end{eqnarray}
\noindent Sum this equation over $j$ to obtain,
\begin{eqnarray}
	\sum_j R_j&\le&\sum_{m=1}^{D_A}\left(\sum_j R(\mu_{m}^j |\overline a_{m}\rangle |b_{m}^j\rangle)\right).
\end{eqnarray}
\noindent Now, $R(\mu_{m}^j |\overline a_{m}\rangle |b_{m}^j\rangle)=1$ if $\mu_m^j \ne 0$ and vanishes otherwise. Then from the last line of the previous paragraph, we have that for each $m$ the quantity in parentheses on the right side of this inequality is less than or equal to $\lfloor D_B/r\rfloor$. This yields,
\begin{eqnarray}
	\sum_j R_j&\le D_A\lfloor D_B/r\rfloor,
\end{eqnarray}
\noindent completing the proof.\hspace{\stretch{1}}$\blacksquare$

We will now see that the bound in the theorem can be reached in all cases not discussed in Section~\ref{rankSum} (that is, whenever that bound does not exceed $D_{\min}N_{\max}$), considering first $D_B \ge D_A$. Then we can have a set of $n_B = \lfloor D_B/r\rfloor$ rank-$D_A$ states, as
\begin{equation}\label{eqn:Schma}
	|\Psi_j\rangle = \sum_{k=0}^{D_A-1} |k\rangle_A|k\oplus_{_B}(j-1)r\rangle_B,~~j = 1,\cdots,n_B,
\end{equation}\label{eqn:Schm1}
with $\oplus_{_{B(A)}}$ here indicating addition mod $D_{B(A)}$. Defining new quantities $n_A$ and $a<r$ through the relation $D_A = n_Ar+a$, we can write Alice's POVM as ($m=1,\cdots,n_A-1$),
\begin{eqnarray}\label{SchmidtPOVM}
	E_m = \sum_{k=0}^{r-1} |k+(m-1)r\rangle_A\langle k+(m-1)r|,\nonumber\\
	E_{n_A} = \sum_{k=0}^{r-1} a_k|k+(n_A-1)r\rangle_A\langle k+(n_A-1)r|,\nonumber\\
	E_{n_A+1} = \sum_{k=0}^{r-1} a_k^\prime |k+(D_A-r)\rangle_A\langle k+(D_A-r)|,	
\end{eqnarray}
where if $a=0$, $a_k=1$ and $E_{n_A+1}$ is to be omitted. When $a\ne0$, $a_k=1/2$ when $k=a\cdots,r-1$ and $a_k^\prime=1/2$ when $k=0,\cdots,r-a-1$; otherwise, these coefficients are equal to one.  Notice that the last two POVM elements have overlapping supports when $a\ne0$, which is why some of the coefficients must differ from unity. Whichever outcome Alice obtains, Bob can distinguish preserving Schmidt rank $r$. An example of such a set of states is given in Fig.~\ref{fig:Schmidt}(a).

If $D_A/n_A\le D_B<D_A$, we cannot have states with Schmidt rank-$D_A$, so instead choose $n_B(n_A-1)$ states having rank $r$ and in addition, $n_B$ states having rank $r+a$, with all these states represented as
\begin{equation}\label{eqn:Schmb}
	|\Psi_{(n-1)n_B+j}\rangle = \sum_{k=0}^{k_n} |k\oplus_{_{A}}(n-1)r\rangle_A|k\oplus_{_{B}}(j-1)r\rangle_B,
\end{equation}
where $n=1,\cdots n_A$, $j = 1,\cdots,n_B$, and $k_n=r-1$ except when $n=n_A$ in which case $k_{n_A}=r+a-1$. Alice does the same POVM given in Eq.~(\ref{SchmidtPOVM}), and as in the previous case for each of her possible outcomes, Bob can distinguish preserving $r$. This set of states is illustrated by the example in Fig.~\ref{fig:Schmidt}(b).

\begin{figure}
\centering
\begin{picture}(160,100)
	\put(-50,12){\makebox(16,20){$|3\rangle_A$}}
	\put(-50,32){\makebox(16,20){$|2\rangle_A$}}
	\put(-50,52){\makebox(16,20){$|1\rangle_A$}}
	\put(-50,72){\makebox(16,20){$|0\rangle_A$}}
	{\thicklines\put(-32,12){\framebox(96,80)}} 
	\put(-32,12){\framebox(16,20){2}\framebox(16,20){}\framebox(16,20){}\framebox(16,20){1}\framebox(16,20){}\framebox(16,20){}} 
	\put(-32,32){\framebox(16,20){}\framebox(16,20){}\framebox(16,20){1}\framebox(16,20){}\framebox(16,20){}\framebox(16,20){2}} 
	\put(-32,52){\framebox(16,20){}\framebox(16,20){1}\framebox(16,20){}\framebox(16,20){}\framebox(16,20){2}\framebox(16,20){}} 
	\put(-32,72){\framebox(16,20){1}\framebox(16,20){}\framebox(16,20){}\framebox(16,20){2}\framebox(16,20){}\framebox(16,20){}} 
	\put(6,-4){\makebox(24,8){(a)}}
	{\thicklines\put(104,12){\framebox(84,90)}} 
	\put(104,12){\framebox(24,10){12}\framebox(24,10){10}\framebox(24,10){11}\framebox(12,10){}} 
	\put(104,22){\framebox(24,20){10}\framebox(24,20){11}\framebox(24,20){12}\framebox(12,20){}} 
	\put(104,42){\framebox(24,20){7}\framebox(24,20){8}\framebox(24,20){9}\framebox(12,20){}} 
	\put(104,62){\framebox(24,20){4}\framebox(24,20){5}\framebox(24,20){6}\framebox(12,20){}} 
	\put(104,82){\framebox(24,20){1}\framebox(24,20){2}\framebox(24,20){3}\framebox(12,20){}} 
	\put(140,-4){\makebox(24,8){(b)}}
\end{picture}
\caption{\label{fig:Schmidt}Examples illustrating how the bound of Theorem~\ref{DADBr} can be achieved. (a) The states of Eq.~(\ref{eqn:Schma}) when $D_A\le D_B$, with $r=3$. Alice's two POVM elements are each rank $3$: the first annihilates $|3\rangle_A$; the second, $|0\rangle_A$. (b) The states of Eq.~(\ref{eqn:Schmb}) when $D_B< D_A$; each square box is $r\times r$. Note that states $10,~11$, and $12$ are rank-($r+a$) with $a\ne0$.}
\end{figure}

\subsection{Preserving the original Schmidt ranks}
\label{keepRj_proof}
$\\$
\noindent {\bf Theorem \ref{KeepRj}} \textit{If a set of states $\{|\Psi_j\rangle\}$ is perfectly distinguishable by LOCC while preserving $R_j$, then the density operators $\{\hat\rho_j\}$ form a mutually orthogonal set.}

\noindent To prove this theorem, we will use the following lemma, in which we refer to a measurement by Alice expressed in terms of Kraus operators expanded as
\begin{eqnarray}\label{kraus}
	A_l = \sum_{m=0}^{D_A-1}|a_m^l\rangle_A\langle m|.
\end{eqnarray}
Then, we can easily prove,
\begin{keepRjLem}\label{KeepRjLem} Given a complete measurement by Alice, the outcomes of which correspond to the operators $A_l$ in Eq.~(\ref{kraus}), and a set of states $|\zeta^l\rangle_A = \sum_m\alpha_m|a_m^l\rangle_A$ with the $\alpha_m$ independent of $l$ and $\alpha_M \ne 0$. Then $\langle a_M^l|\zeta^l\rangle \ne 0$ for at least one outcome of Alice's measurement.
\end{keepRjLem}
Proof: This follows from the fact that for a complete measurement, we have $\sum_lA_l^\dagger A_l = I_A$, or
\begin{equation}
	\sum_l\langle a_m^l|a_{m^\prime}^l\rangle = \delta_{mm^\prime}.
\end{equation}
Then we have that
\begin{equation}
	\sum_l\langle a_M^l|\zeta^l\rangle = \sum_m\alpha_m\sum_l\langle a_M^l|a_{m}^l\rangle = \alpha_M \ne 0.
\end{equation}
The lemma follows directly.\hspace{\stretch{1}}$\blacksquare$

We now prove the theorem.

\noindent Proof of Theorem \ref{KeepRj}: It will be sufficient to show that the parties cannot distinguish a pair of states and preserve $R_j$, $j=1,2,$ if the density operators $\hat\rho_1$ and $\hat\rho_2$ are not orthogonal. Assuming they are not orthogonal, there are two general categories, illustrated in Fig.~\ref{fig:keepRj}, pertaining to the relationship between these two states (Supp($\cdot$) means support of the indicated operator):
\begin{enumerate}
\item Fig.~\ref{fig:keepRj}(a) --- in a product basis expansion of $|\Psi_2\rangle$ there is a term $|\Phi_2^A\rangle|\Phi_2^B\rangle$ such that $|\Phi_2^A\rangle \subseteq \textrm{Supp}(\rho_1^A)$ and $|\Phi_2^B\rangle \subseteq \textrm{Supp}(\rho_1^B)$;
\item Fig.~\ref{fig:keepRj}(b) --- if such a term (as in 1. above) is not in $|\Psi_2\rangle$ then there must be two terms, $|\Phi_2^A\rangle|\Phi_2^B\rangle$ and $|\xi_2^A\rangle|\xi_2^B\rangle$, such that $|\Phi_2^A\rangle \not\perp \textrm{Supp}(\rho_1^A)$ but $|\Phi_2^B\rangle \perp$ Supp($\rho_1^B$), and $|\xi_2^A\rangle \perp$  Supp($\rho_1^A$) while $|\xi_2^B\rangle \not\perp \textrm{Supp}(\rho_1^B)$. 
\end{enumerate}
Let us consider these two cases separately, beginning with the first one. In this case with a convenient choice of bases, the states can be written,
\begin{eqnarray}
	|\Psi_1\rangle &=& \sum_{k,k^\prime=0}^{R_1-1}g_{kk^\prime}|k^\prime\rangle_A|k\rangle_B, \nonumber \\
	|\Psi_2\rangle &=& (\alpha|0\rangle_A + \beta|R_1\rangle_A)|0\rangle_B \nonumber \\
	&+& \sum_{k=1}^{D_B-1}\sum_{k^\prime=0}^{D_A-1}f_{kk^\prime}|k^\prime\rangle_A|k\rangle_B,
\end{eqnarray}	
where the matrix $g$ has rank $R_1$, $\alpha \ne 0$, and we have $|\Phi_2^A\rangle|\Phi_2^B\rangle = \alpha|0\rangle_A|0\rangle_B$ (additional such terms do not change the conclusion).

Then, following outcome $A_l$, written as in Eq.~(\ref{kraus}), we have
\begin{eqnarray}
	A_l|\Psi_1\rangle &=& \sum_{k,k^\prime=0}^{R_1-1}g_{kk^\prime}|a_{k^\prime}^l\rangle_A|k\rangle_B, \nonumber \\
	A_l|\Psi_2\rangle &=& (\alpha|a_0^l\rangle_A + \beta|a_{R_1}^l\rangle_A)|0\rangle_B \nonumber \\
	&+& \sum_{k=1}^{D_B-1}\sum_{k^\prime=0}^{D_A-1}f_{kk^\prime}|a_{k^\prime}^l\rangle_A|k\rangle_B.
\end{eqnarray}	
\noindent Define $|\zeta_0^l\rangle_A = \alpha|a_0^l\rangle_A + \beta|a_{R_1}^l\rangle_A$. There are three possibilities: (a) $|\Psi_1\rangle$ is eliminated by $A_l \Longrightarrow |a_0^l\rangle_A = 0$; (b) $|\Psi_2\rangle$ is eliminated by $A_l \Longrightarrow |\zeta_0^l\rangle_A = 0$; or (c) neither is eliminated by $A_l$, so each must continue to have its original Schmidt rank $\Longrightarrow |a_0^l\rangle_A \ne 0 \ne|\zeta_0^l\rangle_A$. Since $\alpha \ne 0$, we may conclude from Lemma~\ref{KeepRjLem} there must be at least one $l$ such that neither state is eliminated and $\langle a_0^l|\zeta_0^l\rangle \ne 0$. For this outcome, one can again choose a basis with $|\overline 0\rangle_A$ defined as the projection of $|\zeta_0^l\rangle_A$ onto Supp($A_l\rho_1^AA_l^\dagger$) (since $\langle a_0^l|\zeta_0^l\rangle \ne 0$ and $|a_0^l\rangle_A$ is in this support, this projection is guaranteed to be non-zero). Also choose $|\overline R_1\rangle_A$ as the projection of $|\zeta_0^l\rangle_A$ onto the kernel of $A_l\rho_1^AA_l^\dagger$, if this projection is non-zero (otherwise $\overline \beta = 0$ below and the choice of $|\overline R_1\rangle_A$ is unrestricted, the following conclusion being unchanged). This gives
\begin{eqnarray}
	A_l|\Psi_1\rangle &=& \sum_{k,k^\prime=0}^{R_1-1}\overline g_{kk^\prime}|\overline k\rangle_A|k\rangle_B, \nonumber \\
	A_l|\Psi_2\rangle &=& (\overline \alpha|\overline 0\rangle_A + \overline \beta|\overline R_1\rangle_A)|0\rangle_B \nonumber \\
	&+& \sum_{k=1}^{D_B-1}\sum_{k^\prime=0}^{D_A-1}\overline f_{kk^\prime}|\overline k^\prime\rangle_A|k\rangle_B,
\end{eqnarray}	
which has exactly the same form as before Alice's measurement. By the symmetry between the parties, the same conclusion will hold after Bob's subsequent measurement, and by extension, after they complete an arbitrary number of rounds of measurements. In other words, for any LOCC protocol that preserves $R_j$, there will always be an outcome such that they have failed to distinguish between this pair of states.

For the second case, we can choose bases such that
\begin{eqnarray}
	|\Psi_2\rangle &=& \alpha|0\rangle_A|R_1\rangle_B  + \beta|R_1\rangle_A|0\rangle_B\nonumber \\
	&+& \sum_{k=1}^{D_B-1}\sum_{k^\prime=1}^{D_A-1}f_{kk^\prime}|k^\prime\rangle_A|k\rangle_B,
\end{eqnarray}	
with $\alpha \ne 0 \ne \beta$, and after outcome $A_l$,
\begin{eqnarray}
	A_l|\Psi_2\rangle &=& \alpha|a_0^l\rangle_A|R_1\rangle_B  + \beta|a_{R_1}^l\rangle_A|0\rangle_B\nonumber \\
	&+& \sum_{k=1}^{D_B-1}\sum_{k^\prime=1}^{D_A-1}f_{kk^\prime}|a_{k^\prime}^l\rangle_A|k\rangle_B.
\end{eqnarray}	
\noindent First note that if for any single outcome $|a_{R_1}^l\rangle_A$ is not orthogonal to $|a_{k^\prime}^l\rangle_A~\forall_{k^\prime=0,\cdots,R_1-1}$, then we are back to the previous case for which we have seen the parties cannot distinguish the states and preserve $R_j$. So we may assume this orthogonality in the following. Define $|\zeta_{R_1}^l\rangle_A = \alpha|a_0^l\rangle_A + \sum_{k^\prime}f_{R_1k^\prime}|a_{k^\prime}^l\rangle_A$. By an argument similar to that given for the previous case, there must be at least one outcome for which neither state is eliminated and $\langle a_0^l|\zeta_{R_1}^l\rangle \ne 0$. Then choosing a new basis with $|\tilde R_1\rangle_A = |a_{R_1}^l\rangle_A$ and $|\tilde 0\rangle_A$ the projection of $|\zeta_{R_1}^l\rangle$ onto Supp($A_l\rho_1^AA_l^\dagger$), we have
\begin{eqnarray}
		A_l|\Psi_1\rangle &=& \sum_{k,k^\prime=0}^{R_1-1}\tilde g_{kk^\prime}|\tilde k\rangle_A|k\rangle_B, \nonumber \\
A_l|\Psi_2\rangle &=& \tilde\alpha|\tilde0\rangle_A|R_1\rangle_B  + \beta|\tilde R_1\rangle_A|0\rangle_B\nonumber \\
	&+& \sum_{k=1}^{D_B-1}\sum_{k^\prime=1}^{D_A-1}\tilde f_{kk^\prime}|\tilde k^\prime\rangle_A|k\rangle_B.
\end{eqnarray}	
Once again we see that this has the same form as before Alice's measurement, and there is also a symmetry between the parties. Hence by the argument used for the previous case, we must conclude that they cannot distinguish between these states. This covers all possible cases, so the conclusion holds quite generally and the theorem follows directly.\hspace{\stretch{1}}$\blacksquare$

\section{Proof of nonlocality without entanglement}
\label{app:Bennett9}

We give here a very simple and transparent proof that the nine orthogonal product states of Bennett, et.al. \cite{Bennett9}
\begin{eqnarray}
|\Psi_1\rangle & = & |1\rangle_A|1\rangle_B,\nonumber\\
|\Psi_2\rangle & = & |0\rangle_A(|0\rangle_B+|1\rangle_B),\nonumber\\
|\Psi_3\rangle & = & |0\rangle_A(|0\rangle_B-|1\rangle_B),\nonumber\\
|\Psi_4\rangle & = & |2\rangle_A(|1\rangle_B+|2\rangle_B),\nonumber\\
|\Psi_5\rangle & = & |2\rangle_A(|1\rangle_B-|2\rangle_B),\nonumber\\
|\Psi_6\rangle & = & (|1\rangle_A+|2\rangle_A)|0\rangle_B,\nonumber\\
|\Psi_7\rangle & = & (|1\rangle_A-|2\rangle_A)|0\rangle_B,\nonumber\\
|\Psi_8\rangle & = & (|0\rangle_A+|1\rangle_A)|2\rangle_B,\nonumber\\
|\Psi_9\rangle & = & (|0\rangle_A-|1\rangle_A)|2\rangle_B.
\end{eqnarray}
cannot be distinguished by LOCC. The method of proof will be to consider general local operations by either party, and to show that the only ones that do not destroy the mutual orthogonality of the states are proportional to unitaries. Then, since unitary operators do not provide the parties with any information, and also do not alter the relationship between the states, the party who goes next can do no better, and so on, no matter how many rounds of measurements they make. Hence, they are unable to distinguish with certainty. 

Proof: Due to the symmetry between the parties we may suppose Alice goes first, implementing a completely general local operation written as
\begin{equation}
	A = \sum_{k=0}^2 |a_k\rangle_A\langle k|.
\end{equation}
The nine states after this operation become
\begin{eqnarray}
|\Psi_1^\prime\rangle & = & |a_1\rangle_A|1\rangle_B,\nonumber\\
|\Psi_2^\prime\rangle & = & |a_0\rangle_A(|0\rangle_B+|1\rangle_B),\nonumber\\
|\Psi_3^\prime\rangle & = & |a_0\rangle_A(|0\rangle_B-|1\rangle_B),\nonumber\\
|\Psi_4^\prime\rangle & = & |a_2\rangle_A(|1\rangle_B+|2\rangle_B),\nonumber\\
|\Psi_5^\prime\rangle & = & |a_2\rangle_A(|1\rangle_B-|2\rangle_B),\nonumber\\
|\Psi_6^\prime\rangle & = & (|a_1\rangle_A+|a_2\rangle_A)|0\rangle_B,\nonumber\\
|\Psi_7^\prime\rangle & = & (|a_1\rangle_A-|a_2\rangle_A)|0\rangle_B,\nonumber\\
|\Psi_8^\prime\rangle & = & (|a_0\rangle_A+|a_1\rangle_A)|2\rangle_B,\nonumber\\
|\Psi_9^\prime\rangle & = & (|a_0\rangle_A-|a_1\rangle_A)|2\rangle_B.
\end{eqnarray}
We require $\langle\Psi_j^\prime|\Psi_{j^\prime}^\prime\rangle = 0~\forall_{j^\prime \ne j}$ (some of these states may vanish identically), since otherwise they cannot distinguish with certainty. Then, considering in turn $\{j,j^\prime\} = \{1,2\},~\{1,4\}$, and $\{2,4\}$, we conclude that the states $|a_k\rangle_A$ form a mutually orthogonal set (again, some may vanish). Using this fact and considering orthogonality for $\{j,j^\prime\} = \{6,7\}$ and $\{8,9\}$, we see that the $|a_k\rangle_A$ must all have the same norm; $\langle a_k|a_k\rangle$ is independent of $k$. Hence, since we may assume that $A$ does not vanish identically, it must be proportional to a unitary operator, which completes the proof.\hspace{\stretch{1}}$\blacksquare$

\section{Two-way communication is better than one-way}
\label{2vs1}

For Theorem~\ref{DbyN} and Corollary~\ref{Dbyr} (which, with $D_A=D_B=D$ and $R_j=D$, state that $r<D/N$ and $N<D/r$ for one-way communication, respectively), I now show that if they use LOCC-P2, the parties can successfully distinguish which state and also, at least for some outcomes of their measurements, preserve $r = D/2 > D/N$. Let $N=3>D/r = 2$, and let $D \ge 8$ be a multiple of $4$. Take the original states as
\begin{eqnarray}
	\label{threestates}
	|\Psi_1\rangle & = & \sum_{k=0}^{D-1} |k\rangle_A |k\rangle_B, \nonumber \\
	|\Psi_2\rangle & = & \sum_{k=0}^{D-1} |k\rangle_A |k \oplus D/2\rangle_B, \nonumber \\
	|\Psi_3\rangle & = & \sum_{k=0}^{D-1} (-1)^k|k\rangle_A |k \oplus D/2\rangle_B,
\end{eqnarray}
where $\oplus$ denotes addition mod $D$. Alice and Bob can each make an orthogonal measurement on their respective spaces, with outcomes of rank $D/2$ represented by the projectors
\begin{eqnarray}
	P_{\alpha 1} & = & \sum_{k=0}^{D/2-1} |k\rangle_\alpha \langle k|, \nonumber \\
	P_{\alpha 2} & = & \sum_{k=D/2}^{D-1} |k\rangle_\alpha \langle k|.
\end{eqnarray}
One of them, say Alice, tells the other which outcome she obtained. If Bob's outcome was the same as Alice's, then since $P_{Al}\otimes P_{Bl}|\Psi_j\rangle = 0$ for $j=2,3$ but not for $j=1$, he then knows that the state was $|\Psi_1\rangle$ and they now share a state of Schmidt rank $D/2$. If their outcomes were both $l=1$, for example, then they now have
\begin{eqnarray}
	\label{Dby2}
	|\widetilde\Psi_1\rangle = \sum_{k=0}^{D/2-1} |k\rangle_A |k\rangle_B.
\end{eqnarray}

This result is not in contradiction to Theorem~\ref{DbyN} nor to Corollary~\ref{Dbyr}, however, since if their outcomes were not the same, then while the probability that the state was $|\Psi_1\rangle$ now vanishes, the probabilities for the other two states are \textit{both} nonzero. Therefore, in this event one-way communication has been insufficient to distinguish the state. For example, if Alice obtained outcome $1$ while Bob obtained $2$ (the other case works in a similar way), then their system is left in one or the other of 
\begin{eqnarray}
	|\widetilde\Psi_2\rangle & = & \sum_{k=0}^{D/2-1} |k\rangle_A |k + D/2\rangle_B, \nonumber \\
	|\widetilde\Psi_3\rangle & = & \sum_{k=0}^{D/2-1} (-1)^k|k\rangle_A |k + D/2\rangle_B.
\end{eqnarray}
To distinguish between these possibilities, the parties must make another round of measurements. Before Alice measures, she needs to know from Bob whether or not she should. Otherwise, if Bob had obtained the \textit{same} outcome that she obtained, in which case they share the state in Eq.~(\ref{Dby2}), and she goes ahead with the following measurement anyway, they will no longer share a state of Schmidt rank $D/2$. This illustrates the results of Theorem~\ref{DbyN} and Corollary~\ref{Dbyr} in a somewhat nontrivial way.

To complete their task, Bob does a measurement with projectors $\{P_{B+},P_{B-}\}$, where
\begin{eqnarray}
	P_{B\pm} & = & \frac{1}{2}\sum_{k=0}^{D/4-1}\bigl(|2k + D/2\rangle_B \pm |2k + 1 + D/2\rangle_B\bigr)\nonumber \\
		& \times &\bigl(_B\langle 2k + D/2| \pm~_B\langle 2k + 1 + D/2|\bigr).
\end{eqnarray}
Now,
\begin{eqnarray}
	P_{B\pm} |\widetilde\Psi_2\rangle & = & \frac{1}{2}\sum_{k=0}^{D/4-1}\bigl(|2k\rangle_A \pm |2k + 1\rangle_A\bigr)\nonumber \\
		& \times &\bigl(|2k + D/2\rangle_B \pm |2k + 1 + D/2\rangle_B\bigr),\nonumber \\
	P_{B\pm} |\widetilde\Psi_3\rangle & = & \frac{1}{2}\sum_{k=0}^{D/4-1}\bigl(|2k\rangle_A \mp |2k + 1\rangle_A\bigr)\nonumber \\
		& \times &\bigl(|2k + D/2\rangle_B \pm |2k + 1 + D/2\rangle_B\bigr),\nonumber \\
\end{eqnarray}
so Alice can complete the protocol by the measurement $\{P_{A+},P_{A-}\}$, with
\begin{eqnarray}
	P_{A\pm} = \frac{1}{2}\sum_{k=0}^{D/4-1}\bigl(|2k\rangle_A \pm |2k + 1\rangle_A\bigr)
		\bigl(_A\langle 2k| \pm~_A\langle 2k + 1|\bigr).\nonumber\\
\end{eqnarray}
If their outcomes are the same ($++$ or $--$), the state was $|\Psi_2\rangle$ and if different ($+-$ or $-+$) the state was $|\Psi_3\rangle$. In either case, they preserve Schmidt rank of $D/4$.

Another example where two-way communication is better than one-way pertains to Corollary~\ref{Dbyr}, which says that if $N > D/2$, a one-way protocol cannot preserve entanglement for any single outcome while always distinguishing the state.  I now give a case where success is possible for $N > D/2$ with a two-way protocol. In particular, in $D=5$ the set of $N=3$ states --- $|\Psi_0\rangle = \sum_k |k\rangle_A|k\rangle_B$ and $|\Psi_j\rangle = \sum_k |k\rangle_A|k\oplus(1+j)\rangle_B,~j=1,2$ (with $\oplus$ again denoting addition mod $D$) --- may be distinguished with certainty using a two-way protocol, with some outcomes preserving Schmidt rank of $2$. One such successful outcome is when the parties both obtain $P_{\alpha 1} = |0\rangle_\alpha\langle0| + |1\rangle_\alpha\langle1|,~\alpha = A,B$ in their measurements, identifying $|\Psi_0\rangle$ as the state. I leave the remainder of this protocol (which is not unique) as an exercise for the reader. (Here, as in the previous example, communication must go both ways after their initial measurements in order that they know whether or not to continue. Otherwise, they risk destroying entanglement in the case they preserved it, because they do not know the result of the other's measurement.)

\section{Preserving $\lfloor D/N\rfloor$ with LOCC-K0}
\label{cn}

Here, I will prove two theorems for distinguishing and preserving entanglement by LOCC-K0, with the discussion restricted to sets of Schmidt rank-$D$ states in $D\times D$. According to Theorem~\ref{DbyN}, the largest Schmidt rank they can preserve is $D/N$. Let us begin with a theorem concerning the case when $D/N$ is an integer.

\begin{integer}\label{integer} Suppose $D/N$ is an integer, and the parties are restricted to LOCC-K0. If there is a protocol in which they are always able to distinguish a set of Schmidt rank-$D$ states on a $D\times D$ system, and can preserve the maximum Schmidt rank of $D/N$ for at least one outcome, then they preserve $D/N$ for all their outcomes, and they can do so using orthogonal measurements (LOCC-P0).
\end{integer}

\noindent Proof: Consider the single outcome that by assumption preserves $D/N$, for which both Alice's and Bob's measurement operators (say, $A_m,~B_n$) must have rank $D/N$. When Alice gets $A_m$, each of Bob's reduced density operators, $\rho_{jm}^B$, will have rank $D/N$. Then to distinguish with certainty these density operators must all be mutually orthogonal (this is why neither $A_m$ or $B_n$ can have rank greater than $D/N$). Given that $D/N$ is an integer, this uniquely determines an orthogonal measurement Bob may use, with measurement operators having support identical to those of these density operators, so each has rank $D/N$. Hence, when Alice gets $A_m$, they can preserve $D/N$ for any of Bob's outcomes, and he may just as well do an orthogonal measurement. Recognizing that Alice and Bob play completely equivalent roles, this argument may be turned around starting with Bob's outcome $B_n$ determining an orthogonal measurement Alice can choose, thus showing that they both may choose orthogonal measurements for which each of their combined outcomes preserves $D/N$.\hspace{\stretch{1}}$\blacksquare$

It should be noted that the starting assumption that they are always able to distinguish the state is crucial. The restricted assumption, that only the single outcome $\{A_m,B_n\}$ is known to distinguish the state, certainly does not lead to the conclusion that the parties can distinguish for all outcomes. This does not even follow from knowing that the single outcome on Alice's side $A_m$ is known to allow Bob to distinguish for every outcome of a measurement he can make, a fact that can be seen from the following discussion. As argued in the above proof, $A_m$ determines a set of reduced density operators on Bob's side, which determine the allowed supports for Bob's measurement operators, which in turn determine a set of allowed supports for Alice's measurement operators. There is certainly no guarantee that these \textit{sets} of supports, one for each of Bob's outcomes, will all be identical. If they are not, Alice will be unable to choose a measurement that will distinguish in all cases, and the parties must fail for at least some of their outcomes.

What if $D/N$ is not an integer, but it is known they can always distinguish and that one outcome preserves Schmidt rank of $\lfloor D/N\rfloor $? In this case, the extra dimensions allow for flexibility in the choice of measurements, and the conclusion of the previous theorem no longer holds. Indeed, the two states ($D = 5$, $N=2$)
\begin{eqnarray}
	|\Psi_1\rangle & = & \sum_{k=0}^4 |k\rangle_A |k\rangle_B, \nonumber \\
	|\Psi_2\rangle & = & |0\rangle_A |2\rangle_B + |1\rangle_A |3\rangle_B \nonumber \\
		& + & |2\rangle_A |0\rangle_B + |3\rangle_A |4\rangle_B + |4\rangle_A |1\rangle_B,
\end{eqnarray}
\noindent are distinguished for all outcomes by the orthogonal measurements
\begin{eqnarray}
	P_{\alpha 1} & = & |0\rangle_\alpha\langle 0| + |1\rangle_\alpha\langle 1|, \nonumber \\
	P_{\alpha 2} & = & |2\rangle_\alpha\langle 2| + |3\rangle_\alpha\langle 3|, \nonumber \\
	P_{\alpha 3} & = & |4\rangle_\alpha\langle 4|,
\end{eqnarray}
\noindent but only some outcomes preserve $r = 2$ (outcome $1$ for Alice, $1$ for Bob, for example) whereas others preserve $r = 1$ ($2$ for Alice, $1$ for Bob). While in this example they do not always preserve entanglement, one can easily think up other such examples with $N \ge 3$ where they do.

The following result, which is clearly weaker than Theorem~\ref{integer}, applies to the case when $D/N$ is not an integer.

\begin{nonint} \label{nonint}Suppose $D/N$ is not an integer, and the parties are restricted to LOCC-K0. Then in order for them to distinguish a set of Schmidt rank-$D$ states on a $D\times D$ system and preserve the maximum Schmidt rank of $\lfloor D/N\rfloor$ for every outcome of their measurement, it must be that $D = (N + n)\lfloor D/N\rfloor$, with $n$ a positive integer. That is, it must be possible to divide the parties' spaces into subspaces all having dimension equal to the maximum achievable Schmidt rank, a task for which they can use LOCC-P0.
\end{nonint}

\noindent Proof: First note that if all outcomes preserve $\lfloor D/N\rfloor$, each of the parties' measurement operators must have rank $\lfloor D/N\rfloor$ (this follows from arguments similar to those in the preceding proof). I will show below that any pair of measurement operators for either one of the parties must have supports that are either orthogonal or identical. Then, the measurements divide their spaces into orthogonal subspaces each of dimension $\lfloor D/N\rfloor$, and the theorem follows immediately. 

The proof is by contradiction. Hence, suppose on the contrary $A_1$ and $A_2$ have supports of dimension $\lfloor D/N\rfloor$ that are neither orthogonal nor identical to each other. Each ${A}_m~(m=1,2)$ determines a set of $N$ reduced density operators on Bob's space, $\rho_{jm}^B$, one for each state $|\Psi_j\rangle$. Each of these density operators has rank $\lfloor D/N\rfloor$ or $0$. In order to distinguish and preserve $\lfloor D/N\rfloor$, Bob must choose each of his measurement operators to have support containing the support of one of the $\rho_{jm}^B$ and orthogonal to the others (for each $m$). Consider the density operators $\rho_{J1}^B~(\rho_{J2}^B)$ for Alice's first two outcomes and some fixed state $|\Psi_J\rangle$, and Bob's corresponding measurement operators ${B}_1~({B}_2)$. The support of the latter must be chosen to contain the range of ${M}_J{A}_1~({M}_J{A}_2)$ ($M_J$ is the matrix corresponding to $|\Psi_J\rangle$ in Eq.~(\ref{ensemble})). Given $|\Psi_J\rangle$ is rank-$D$, then ${M}_J$ is non-singular, and the fact that the supports of ${A}_1$ and ${A}_2$ are neither orthogonal nor identical implies the same fact about the supports of ${B}_1$ and ${B}_2$. In particular, the range of ${M}_J{A}_1$, which is the support of $\rho_{J1}^B$, intersects both the support and the kernel of ${B}_2$. This implies that when the state is $|\Psi_J\rangle$ and Alice measures ${A}_1$, there is a non-zero probability that Bob will measure ${B}_2$ (since the support of ${B}_2$ is not orthogonal to $\rho_{J1}^B$). When this occurs, Bob's reduced density operator becomes $\widetilde\rho_{J1}^B = {B}_2\rho_{J1}^B{B}_2^\dagger$ (ignoring unimportant normalization). But given that the support of $\rho_{J1}^B$ is not orthogonal to the kernel of ${B}_2$, the rank of $\widetilde\rho_{J1}^B$ will be strictly less than that of $\rho_{J1}^B$. That is, the Schmidt rank of their residual shared state, which is equal to the rank of $\widetilde\rho_{J1}^B$, is strictly less than $\lfloor D/N \rfloor$, the rank of $\rho_{J1}^B$. This contradicts the conditions of the theorem, implying that ${A}_1$ and ${A}_2$ must have supports that are either orthogonal or identical to each other.\hspace{\stretch{1}}$\blacksquare$

Theorem \ref{nonint} only gives a necessary condition, so it says nothing as to whether or not the ability to divide into equal size subspaces of dimension $\lfloor D/N\rfloor$ is sufficient for the parties to accomplish this task. The following example shows that when $D = (N + n)\lfloor D/N\rfloor$, there exists at least one set of $N$ states that can always be distinguished by LOCC-P0 preserving the maximal possible Schmidt rank. The states are
\begin{equation}
	\label{shift}
	|\Psi_j\rangle = \sum_{k=0}^D |k\rangle_A |k \oplus \lfloor D/N\rfloor (j-1)\rangle_B,~j=1,\cdots,N
\end{equation}
with $\oplus$ again denoting addition mod$~D$. The parties do the orthogonal measurements,
\begin{equation}
	\label{Ps}
	P_{\alpha l} = \sum_{k=\lfloor D/N\rfloor (l-1)}^{\lfloor D/N\rfloor l-1} |k\rangle_\alpha\langle k|,~l=1,\cdots,N+n.
\end{equation}
Then, $P_{A l} \otimes P_{Bl^\prime} |\Psi_j\rangle$ vanishes except \cite{explain} for the single state $j$ satisfying $l^\prime = (j + l - 1)~\textrm{mod}~(N+n)$, and since $1 \le j \le N$, then for fixed $l$ the possible set of values for $l^\prime$ are $l,l+1,\cdots,(l+N-1)~\textrm{mod}~(N+n)$. When it does not vanish, it is equal to
\begin{equation}
	P_{A l} \otimes P_{Bl^\prime} |\Psi_j\rangle  =  \sum_{k=\lfloor D/N\rfloor l}^{\lfloor D/N\rfloor (l+1)-1} |k\rangle_A |k \oplus \lfloor D/N\rfloor (j-1)\rangle_B,
\end{equation}
which is of Schmidt rank $\lfloor D/N\rfloor $. Thus, all outcomes with non-zero probability distinguish, preserving $\lfloor D/N\rfloor$.


\begin{thebibliography}{10}
\bibitem{twined}
The word ``intertwined" is perhaps not ideal here, but ``entangled" has already
  been spoken for.
\bibitem{Th1stronger}
A bound involving $N$ and $\overline R$ for two-way communication can also be
  obtained by a straightforward alteration of the proof for Theorem~\ref{DBy2},
  but it is always strictly weaker than the one given in that theorem.
\bibitem{explain}
This can be seen by noting that the measurement operators
  $P_{\alpha l}$ divide the respective spaces into $N+n$ equal size subspaces,
  and the states $|\Psi_j\rangle$ are isomorphic (see \cite{ZyczBeng} and
  references therein) to maps that take the $l^{th}$ of Alice's subspaces to
  the $[l+j-1~\textrm{mod}(N+n)]^{th}$ of Bob's. Hence for fixed $j,l$, the
  single integer $l^\prime = l+j-1~\textrm{mod}(N+n)$ corresponds to the only
  one of Bob's outcomes with nonzero probability.
\end{thebibliography}

\newpage

\end{document}